\documentclass{emulateapj}
\usepackage{lscape}
\newcommand{\tq}{t_\mathrm{q}}

\begin{document}

\shorttitle{Stellar Populations of Spirals in Virgo}
\shortauthors{Crowl \& Kenney}

\title{The Stellar Populations of Stripped Spiral Galaxies in the
  Virgo Cluster}

\author{Hugh H. Crowl\altaffilmark{1}}
\author{Jeffrey D.P. Kenney}

\affil{Department of Astronomy, Yale University, New Haven, CT 06520}

\altaffiltext{1}{Current Address: Department of Astronomy, University of
  Massachusetts, 710 North Pleasant Street, Amherst, MA 01003-9305; hugh@astro.umass.edu}

\begin{abstract}

We present an analysis of the stellar populations of the gas-stripped
outer disks of ten Virgo Cluster spiral galaxies, utilizing SparsePak
integral field spectroscopy on the WIYN 3.5m telescope and GALEX UV
photometry. The galaxies in our sample show evidence for being
gas-stripped spiral galaxies, with star formation within a truncation
radius, and a passive population beyond the truncation radius. We find
that all of the galaxies with spatially truncated star formation have
outer disk stellar populations consistent with star formation ending
within the last 500~Myr. The synthesis of optical spectroscopy and
GALEX observations demonstrate that star formation was relatively
constant until the quenching time, after which the galaxies passively
evolved.  Large starbursts at the time of quenching are excluded for
all galaxies, but there is evidence of a modest starburst in at least
one galaxy. For approximately half of our galaxies, the timescales
derived from our observations are consistent with galaxies being
stripped in or near the cluster core, where simple ram-pressure
estimates can explain the observed stripping. However, the other half
of our sample galaxies were clearly stripped outside the cluster
core. Such galaxies provide evidence that the intra-cluster medium is
not static and smooth. For three of the most recently stripped
galaxies, there are estimates for the stripping timescales from
detailed gas stripping simulations. For all three of these galaxies,
our stripping timescales agree with those from the gas stripping
simulations, suggesting that star formation is quenched near the time
of peak pressure. While the stripping of star-forming gas in the outer
disk creates a passive population in our galaxies, there is still
normal star formation in the center of our sample galaxies.  It may be
that Virgo is not massive enough to completely strip these spiral
galaxies and, in a more dynamically active cluster or a cluster with a
higher density ICM, such a process would lead to passive spirals
and/or S0s.

\end{abstract}


\section{Introduction}

The morphology-density relationship \citep{dressler80} implies that
clusters are creating S0 galaxies at the expense of spirals.  The most
salient difference between the two galaxy populations is the presence
(spirals) or absence (S0's) of current strong star formation.
\citet{kauffmann04} find that the star formation - environment
correlation is the strongest correlation of any galaxy property with
environment, suggesting that environment plays a key role in the
termination of star formation.  While there is intense debate as to
the specific environmental processes that cause the transformation,
stellar population simulations \citep{shioya04} have shown that simple
quenching of star formation in a spiral galaxy can form S0-type
spectra. However, not all S0 galaxies are faded spirals; there is a
significant fraction of S0's with larger bulges than spiral galaxies
(e.g.  \citealp{christlein04}), something that cannot be explained by
simple disk fading. Ram pressure stripping is not the sole cause of
the transformation of spirals to S0's, but it does appear to be a key
part of the transformation for many galaxies.

A recently discovered population of cluster galaxies, ``passive
spirals'' \citep{dressler99}, have noticeable spiral structure but
little, if any, ongoing star formation.  UV photometry \citep{moran06}
has shown that these galaxies have intermediate stellar population
properties between spirals and S0's; passive spirals have no ongoing
star formation, but have stopped forming stars only recently. These
galaxies are consistent with an intermediate phase between spirals and
S0s. \citet{poggianti06} suggest that S0's may represent a
heterogeneous population: ``primordial passive galaxies'' that formed
at $z>2.5$ and environmentally-formed S0s, which may evolve through a
passive spiral phase. \citet{shioya04} demonstrate that such evolution
from active spiral to passive spiral to S0 is naturally explained by
passive stellar evolution after a stripping event and also suggest
that there are at least two paths to S0 galaxies.

In the nearby Virgo cluster, there exists a population of spiral
galaxies with normal star formation inside some radius, but little or
no star formation beyond that ``truncation radius'' \citep{kk04}.
Many of these galaxies have relatively undisturbed stellar
distributions, despite their severely truncated H$\alpha$ disks. Such
a disturbance in the gas with little or no apparent stellar disk
disturbance strongly suggests a gas-gas interaction; if a
gravitational interaction disturbed the star formation, the stellar
light should also be disturbed. The appearance of these galaxies
beyond the stripping radius (i.e. normal stellar disk with no current
star formation) is strikingly similar to the passive spirals in higher
redshift clusters. By studying the stellar populations of our sample
galaxies, we aim to understand when and where in the cluster galaxies
are stripped. Are these Virgo spirals with truncated star-forming
disks just passive spirals that have not been fully stripped?

Our sample of galaxies are drawn nearly entirely from the work of
\citet{koopmann01}, an atlas of $R$-band and H$\alpha$-band images of
spiral galaxies in Virgo. They find that $\sim 50\%$ of spiral
galaxies in Virgo have truncated H$\alpha$ disks
\citep{kk04}: relatively normal star formation in their central
regions, with little or no star formation beyond a well-defined radius
which we call the gas truncation radius. By studying the stellar
populations of the outer disks of these galaxies through optical
spectroscopy and UV imaging, we aim to understand where in clusters
galaxies can be stripped and what effect gas stripping has on the star
formation properties of spiral galaxies. The location in the cluster
of stripping tells us how effective stripping is at different
distances from the cluster center. Simple models of ISM-ICM stripping
suggest that stripping can only occur in the cluster core, where the
ICM densities are highest. Galaxies stripped far from the center of
the cluster suggest a non-static, non-smooth ICM or extreme conditions
outside the cluster core.

In \S \ref{sec-obs}, we outline our observations and explain the
observational setup for optical spectroscopy and GALEX UV imaging. In
\S\ref{sec-abslines}, we discuss the relevant spectral features that
we use to determine the stellar population properties of these
galaxies. In \S\ref{sec-models}, we describe the models that we have
used to interpret the stellar population properties of our observed
sample. In \S \ref{sec-galaxies}, we detail our observations and
results and discuss what the results imply about the quenching of
star formation in the galaxies' outer disks. In \S
\ref{sec-discussion}, we discuss the implications of our individual
galaxy results for the cluster as a whole, compare our results with
simulations of neutral gas, and briefly discuss what our observations
of these galaxies in Virgo may imply for galaxy evolution in
clusters. Finally, in \S \ref{sec-summary}, we summarize our results.

\section{Observations}
\label{sec-obs}
\subsection{Optical Spectroscopy}

Galaxies in the sample were observed the nights of March 28, 2003 -
April 1, 2003 and March 18 - 23, 2004 on the WIYN 3.5m telescope. For
both runs, we used an identical spectrographic setup: the Bench
Spectrographic camera with the 600@10.1 grating, tuned to provide
spectral coverage between 3900 \AA~and 6800 \AA~at a sampling of 1.4
\AA/pixel and a resolution of $\sim 5.5$ \AA~(FWHM). We used the
SparsePak Formatted Field fiber instrument \citep{bershady04} in order
to get large spatial coverage (80\arcsec x 80\arcsec) of the Virgo
galaxies. While the spectral window extends from quite blue ($\sim
3900$\AA) to fairly red ($\sim 6800$ \AA) the observing efficiency is
not uniform over the entire range: fiber transmission causes
significantly reduced throughput in the blue part of the spectrum and
the grating sensitivity causes a drop in throughput in the red.

\begin{figure*}[p]

    \plotone{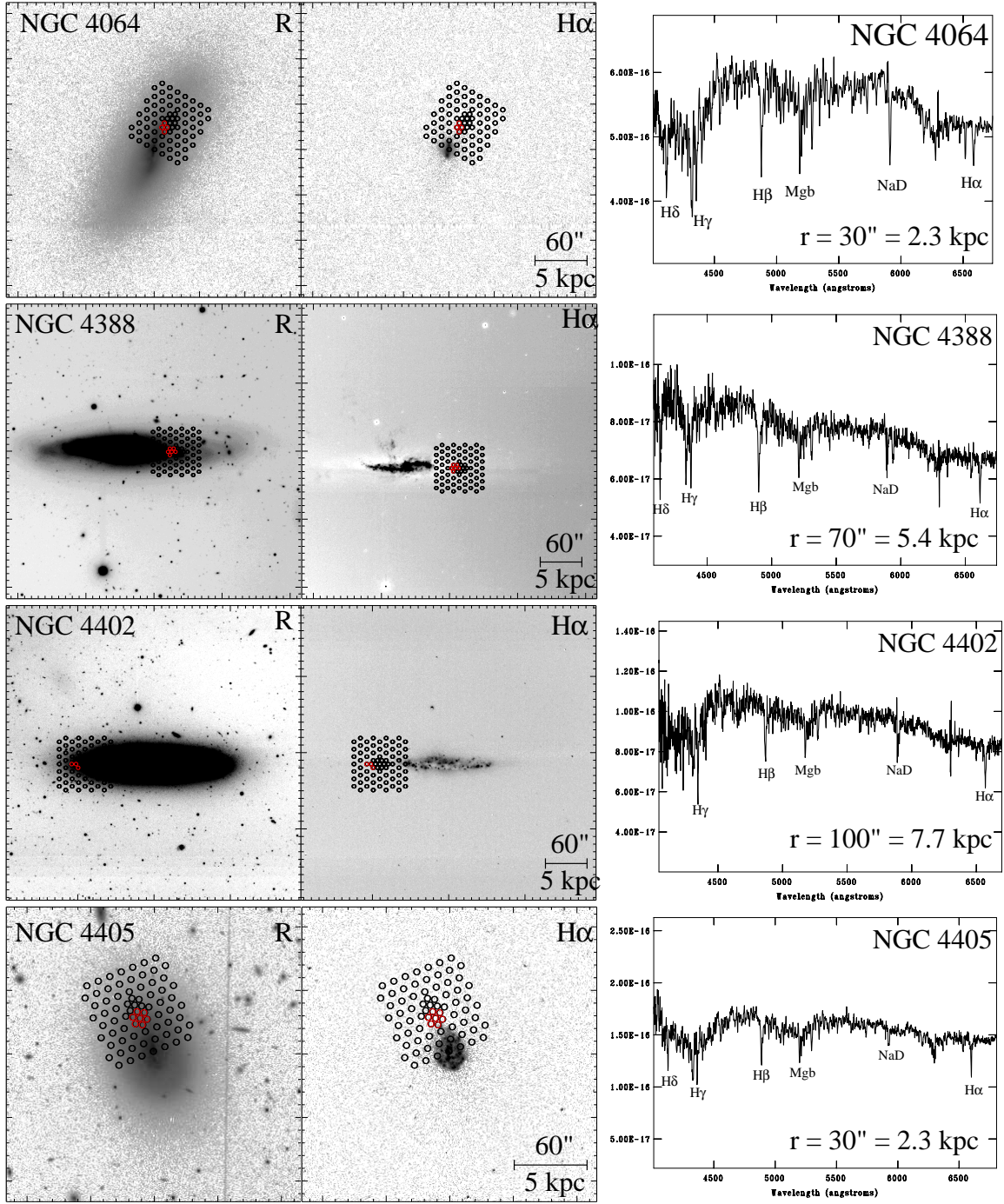}
    \caption[Galaxy SparsePak Positions \& Spectrum]{SparsePak
      Positions on R-band image (left) and H$\alpha$ image
      (center). The composite spectrum from several summed fibers (indicated by
      the red circles on the images) is also shown (right). The radius
      given for each composite spectrum is the distance from the
      galaxy center to the center of the composite spectrum region. Shown here
      are images and spectra for NGC~4064, NGC~4388, NGC~4402, and NGC~4405.}
    \label{fig-gp1sppos}
  \end{figure*}

\begin{figure*}
   \plotone{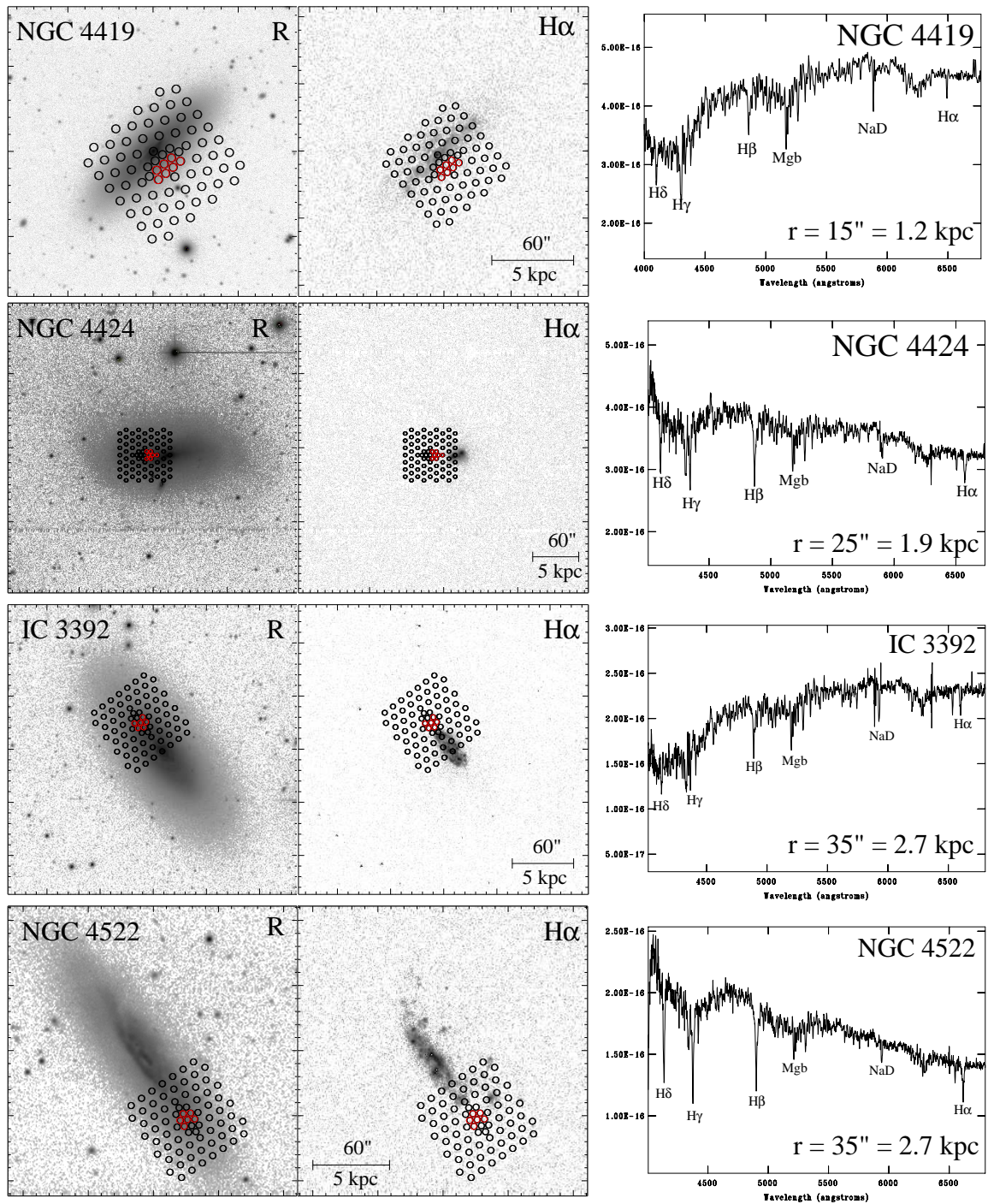}
    \caption{Same as Figure \ref{fig-gp1sppos}, but for NGC~4419,
    NGC~4424, IC~3392, and NGC~4522. Note that the radial distance
    given for NGC~4419 is distance along the {\it minor} axis from
    the galaxy's center.}
    \label{fig-gp2sppos}
  \end{figure*}

\begin{figure*}
       \plotone{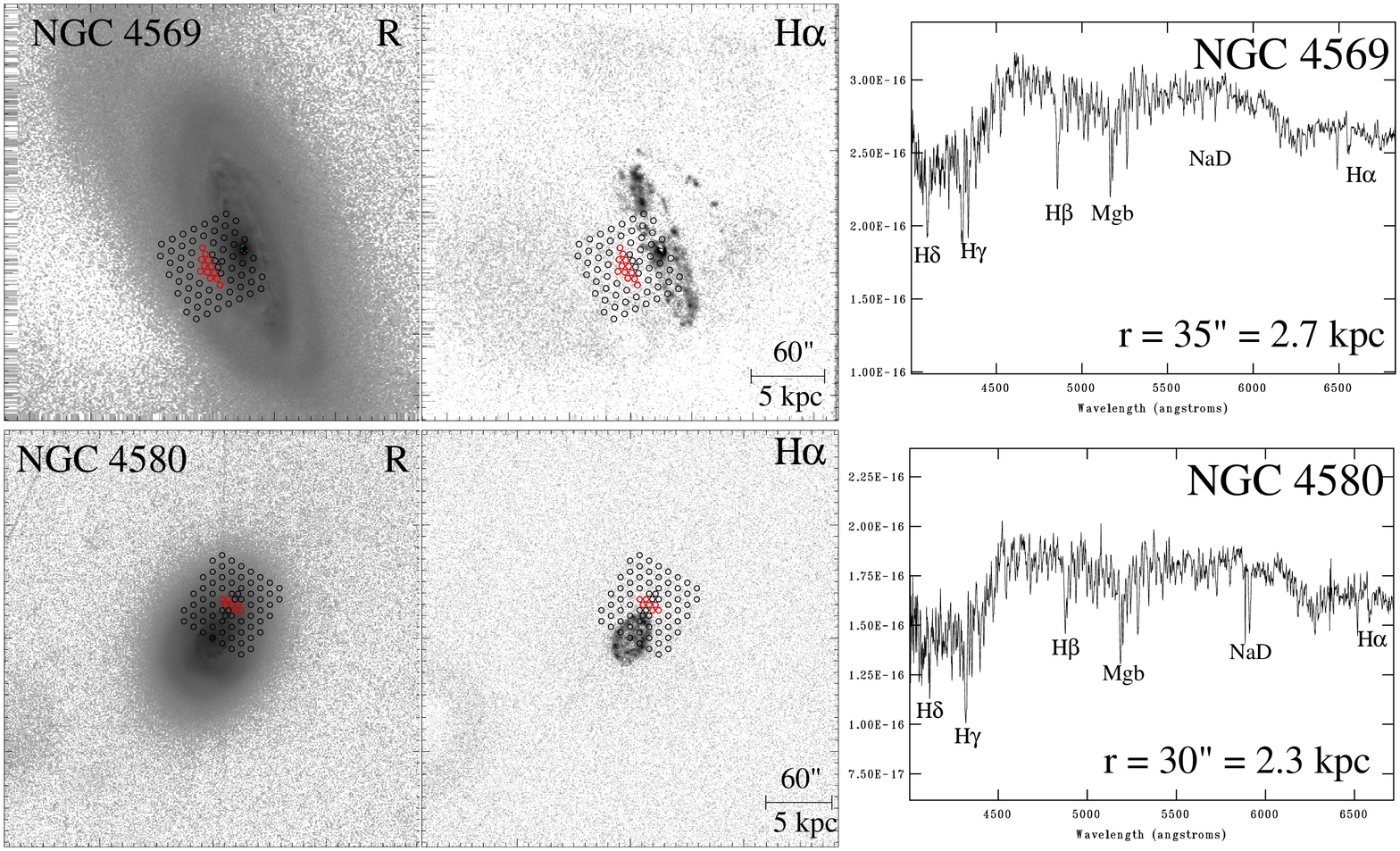}
    \caption{Same as Figure \ref{fig-gp1sppos}, but for NGC~4569 and
    NGC~4580. Note that the radial distance given for NGC~4569 is
    distance along the {\it minor} axis from the galaxy's center.}
\label{fig-gp3sppos}
  \end{figure*}

Several factors were considered when deciding an observational
strategy for each of the disk galaxies in our sample. First, we must
align the SparsePak array on the galaxy such that it's outside the
region of significant H$\alpha$ emission. This part of the galaxy,
beyond the region of ongoing star formation, is interesting because it
is a stellar clock telling us when star formation was interrupted in
that region of the disk. Moreover, emission from star-forming regions
would fill in the age-sensitive Balmer absorption lines, leading to
artificially old age estimates and complicating
interpretations. Observationally, this makes
it possible to observe the stellar populations in the disks of these
galaxies by placing a majority of the SparsePak array outside the
region of star formation (as shown in Figures \ref{fig-gp1sppos} -
\ref{fig-gp3sppos}).
  
  As an additional observational constraint, we observed as close to
  the H$\alpha$ truncation radius as possible. This was done for both
  practical and scientific reasons. From an observational standpoint,
  the regions of the stellar disks are low surface brightness (often,
  at the H$\alpha$ truncation radius, $\mu_R > 21$ mag/arcsec$^2$), so
  it is necessary to observe in the highest surface brightness
  regions.  From a scientific perspective, we are interested in when
  the galaxy was stripped of its star-forming gas. The inner regions
  should be the final regions to be stripped in an interaction with
  the ICM.  Therefore, by studying the region of the galaxy close the
  stripping radius, we can determine how long ago the galaxy
  experienced strong pressure. If the galaxy is past the time of peak
  pressure stripping, this timescale will tell us how long ago peak
  pressure occurred.
  
  The SparsePak array was positioned such that the highest density of
  fibers is nearest to the truncation radius. In most cases, the array
  was placed such that the central ``diamond'' of fibers was arranged
  along the major axis, just beyond the truncation radius. In some
  instances, however, the fiber array was placed parallel to the minor
  axis (i.e. NGC~4419, NGC~4569). In the case of NGC~4419, there is
  extended, low-level H$\alpha$ emission throughout the main disk, so
  we placed the array to attempt to avoid some of that emission. In
  the case of NGC~4569, we chose the location to ensure that we could
  effectively subtract sky light from the galaxy spectra. One
  potential complication associated with observing along the minor
  axis is that bulge stars may contaminate the light from the stellar
  disk. However, both of these galaxies have rather small bulges and
  in neither case do we use fibers placed on the center of the galaxy
  for our analysis. Therefore, we do not expect bulge light to
  significantly affect our results.

  The observations were reduced using the standard IRAF procedures,
  including zero correction and flat field correction. Following basic
  image calibration, spectra were extracted using the {\tt dohydra}
  task in IRAF, as well as wavelength calibrated and sky subtracted
  (using several SparsePak sky fibers). The data were flux calibrated
  using the spectrophotometric standard Feige~34. Finally, individual
  exposures are combined to create a set of 75 spectra for each
  telescope pointing, with each spectrum being from one fiber
  ($4\farcs7$ or 360 pc\footnote{Here, and throughout this paper, we
    assume a distance to Virgo of 16 Mpc.} in diameter).

  \begin{deluxetable}{l|c c c}
    \tablewidth{0pt}
\tablehead{\colhead{Galaxy} & \colhead{SparsePak $t_\mathrm{exp}$} & 
  \colhead{PA$_{\mathrm{SP}}$} & \colhead{GALEX $t_\mathrm{exp}$}}
     \tablecaption{Optical Spectroscopy and GALEX Observation Log}
\startdata
      NGC~4064 & 3h45m & 10\degr & \nodata  \\\hline
      NGC~4388 & 3h45m & 270\degr & 1511s \\\hline
      NGC~4402 & 3h45m & 90\degr & 1511s  \\\hline
      NGC~4405 & 3h45m & 22\degr & 2174s  \\\hline
      NGC~4419 & 3h & 310\degr & 1061s  \\\hline
      NGC~4424 & 3h45m & 90\degr & 2017s  \\\hline
      IC~3392 & 3h45m & 37\degr & 1702s  \\\hline
      NGC~4522 & 3h45m & 215\degr  & 2505s \\\hline
      NGC~4569 & 3h30m & 205\degr & 1445s\\ \hline
      NGC~4580 & 3h45m & 330\degr & 1453s
      \enddata
  \end{deluxetable}
  
\begin{figure}
       \plotone{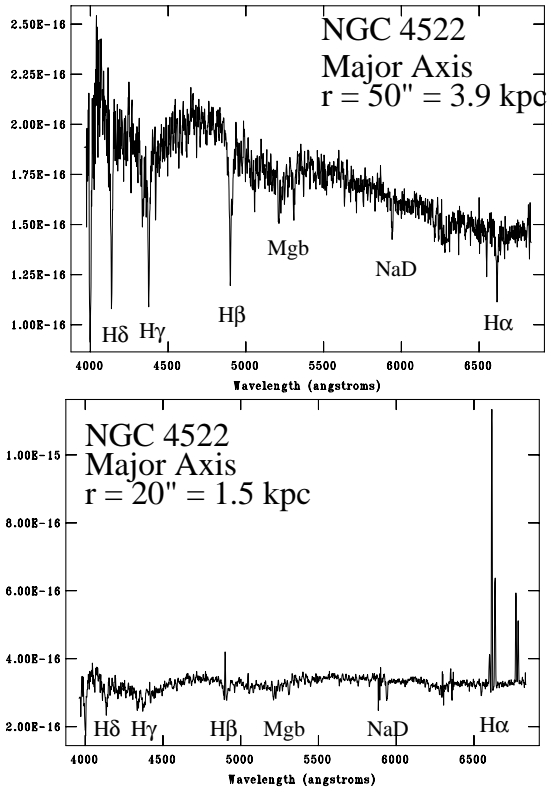}
       \caption[NGC 4522: Two Spectra]{Spectra from the disk of two
         regions of NGC~4522. The top spectrum is from Fiber 52, in
         the disk beyond the truncation radius (50$\arcsec$ from the
         center and 10$\arcsec$ from the H$\alpha$ truncation
         radius). The bottom spectrum is from Fiber 53, closest to the
         center of the galaxy and still in the H$\alpha$ emitting
         region. Note that there is no noticeable emission in the top
         spectrum and that H$\alpha$ is seen in absorption.}
       \label{ngc4522-speccomp}
     \end{figure}

  The spectra are then examined, fiber by fiber, for any residual
  emission that may be present in the regions of the galaxy observed.
  In most cases, it appears the H$\alpha$ images \citep{koopmann01}
  faithfully trace the emission and the spectra show little, if any,
  residual H$\alpha$ emission beyond the truncation radius defined in
  the images. A comparison of two spectra from different regions of
  NGC~4522 shown in Figure \ref{ngc4522-speccomp} demonstrates the
  large differences between a spectrum taken beyond the truncation
  radius (top spectrum) and a spectrum in the star-forming disk
  (bottom spectrum). The metal lines (Mgb and NaD) are largely
  unaffected, but the lower-order Balmer lines are clearly seen in
  emission in the star-forming disk; in particular, H$\beta$ is filled
  in by star-forming emission.  While this effect decreases markedly
  for the higher order lines (\citealp{osterbrock89} reports emission
  line ratios of H$\alpha$/H$\beta$ = 2.86, H$\alpha$/H$\gamma$ =
  6.09, H$\alpha$/H$\delta$ =10.92 for a $10^4$K emission region),
  these lines still suffer from emission fill-in in the inner
  disk. While we went to some effort to exclude fibers with H$\alpha$
  emission from the composite spectra that we analyzed, it became
  apparent through the analysis that there was some minor residual
  Balmer emission in the outer disk in nearly all of the galaxies. The
  contamination is minor, typically less than $0.25$ \AA~for H$\beta$
  and much lower for H$\gamma$ and H$\delta$. Several fibers beyond
  the truncation radius are combined for each galaxy to increase the
  signal to noise ratio; these composite spectra are used for all
  forthcoming spectral analysis.

    For our data, which has intermediate spectral resolution and
    intermediate signal-to-noise ratio, we have chosen to use the
    Lick/IDS spectral indices \citep{faber85}. These indices are
    extracted from data by defining a spectral bandpass and two
    pseudo-continuum bandpasses on either side of the feature of
    interest.

    We chose to extract Lick indices as opposed to directly measuring
absorption lines because of the objectivity of the Lick system; while
fitting absorption lines is appealing for very strong features, the
Lick system allows us to objectively extract spectral features without
subjectively choosing which features of a line to fit. It should be
noted that the indices that we extracted are {\it not} on the
Lick/IDS system, since that would require smoothing our data to even
lower spectral resolution. Instead, we leave our data at its native
spectral resolution and just use the Lick index {\it definitions}
\citep{worthey94b,worthey97} to compare our data with models.

In practice, the spectra are analyzed using the the LECTOR task, which
has been made publicly available by A. Vazdekis\footnote{See
http://www.iac.es/galeria/vazdekis/vazdekis\_software.html}. This
Fortran code numerically integrates an input spectrum for several
indices using Lick index definitions. Additionally, using S/N
estimates of the spectrum, LECTOR measures the uncertainty in an
index. The uncertainty in spectral indices that we quote throughout
this work are the errors due to variations in the S/N and do not take
into account any potential (unknown) calibration errors.

\subsection{GALEX UV Imaging}

For nine of our ten galaxies, we have also utilized GALEX imaging, from
the 100s exposure All-Sky Imaging Survey (AIS) and from the $\sim$
1000s exposure Nearby Galaxy Survey (NGS) \citep{martin05}. Moreover,
we also have included, where possible, data from our medium-depth
($1500 - 2500$ s) Virgo GALEX survey, which gives complete
medium-depth coverage to the galaxies in the VIVA sample (Chung et al
2008, in prep).  The data include both far-UV (FUV;
$\lambda_\mathrm{eff} = 1516$ \AA) and near-UV (NUV;
$\lambda_\mathrm{eff} = 2267$ \AA) images and, in all cases, the
galaxy was detected in the NUV.

From this GALEX imaging, we extract fluxes in circular apertures that
match the size and location of the optical spectroscopy fibers. The
appropriate apertures are then summed to determine the UV flux from
the region of the composite spectrum. The FUV/NUV flux ratio for each
galaxy is a further constraint on the stellar population; because of
how fast the FUV luminosity fades (see Figure \ref{fig-uvsed}), this
measurement is extraordinarily sensitive to recent star formation for
those galaxies where we are able to detect FUV emission. For the
galaxies where FUV emission is undetected, the ratio of the FUV upper
limits \citep{morrissey05} to the NUV flux gives an upper limit to the
stellar population age. We have corrected all GALEX fluxes for
Galactic absorption, using the \citet{schlegel98} reddening maps and
the extinction law of \citet{li01}.

Due to uncertainties with pointing of ground-based spectroscopy we
have gone to some care to make certain that we are sampling the same
region with GALEX photometry as we are with SparsePak spectroscopy. We
have verified that the highest optical flux is present in the fiber
centered at the nucleus of the galaxy in the SparsePak
observations. Moreover, we have verified that strong H$\alpha$
emission is not present in fibers that appear to be beyond the
star-forming region in our images (Figures \ref{fig-gp1sppos} - \ref{fig-gp3sppos}). In
addition, we have verified that even sizable positional offsets do not
significantly impact the color ratios. We have found that the flux
ratios typically change by less than $\pm0.15$ (0.05 in the log) for
offsets up to $\sim 2.3 \arcsec$ (the aperture
radius). Moreover, in several cases, the colors are robust to
significantly larger offsets.

\begin{deluxetable*}{l r c c}
  \tablewidth{0pt}
  \tablecolumns{4}
  \tablecaption{GALEX Photometric Results}
  \tablehead{
    \colhead{Galaxy} &
    \colhead{$F_\mathrm{FUV}$} &
    \colhead{$F_\mathrm{NUV}$} &
        \colhead{$\log (F_\mathrm{FUV}/F_\mathrm{NUV})$}\\
        & \multicolumn{2}{c}{\rule[1mm]{7mm}{0.1mm} (erg/s/cm$^2$/\AA)
  \rule[1mm]{7mm}{0.1mm}} & 
        }
    \startdata
    NGC~4388 & $1.12\pm 0.12 \times 10^{-16}$ & $1.59 \pm 0.05 \times
  10^{-16}$ & $-0.15 \pm 0.11$ \\
    NGC~4402 & $4.65\pm 0.73 \times 10^{-17}$ & $5.74 \pm 0.32 \times
  10^{-17}$ & $-0.09 \pm 0.16$\\
    NGC~4405 & $6.03 \pm 0.68\times 10^{-17}$ & $1.36 \pm 0.04 \times
  10^{-16}$ & $ -0.35 \pm 0.11$\\
    NGC~4419 & $<1.7 \times 10^{-16}$ & $2.74 \pm 0.08 \times 10^{-16}$ & $< -0.31$\\
    NGC~4424 & $1.79 \pm 0.12 \times 10^{-16}$ & $3.49 \pm 0.07 \times 10^{-16}$ &
    $-0.29 \pm 0.07$\\
    IC~3392 & $<2.8 \times 10^{-17}$ & $8.28 \pm 0.37 \times 10^{-17}$ & $<-0.48$\\
    NGC~4522 & $5.18 \pm 0.18 \times 10^{-16}$ & $4.28 \pm 0.06 \times
  10^{-16}$ & $0.08 \pm 0.04 $\\
    NGC~4569 & $1.83 \pm 0.15 \times 10^{-16}$ & $3.99 \pm 0.09 \times 10^{-16}$ &
    $-0.34 \pm 0.09 $\\
    NGC~4580 & $<2.4 \times 10^{-17}$ & $7.19 \pm 0.35 \times
  10^{-17}$ & $< -0.47$\\
  \enddata
\label{tab-galex}  
\end{deluxetable*}

\section{Optical Absorption Line Diagnostics}
\label{sec-abslines}
\subsection{Age-Sensitive Indices}

The optical spectra that we use for our analysis are representative of
the luminosity-weighted mean population for the region of the galaxy
they are extracted from. This means that the observed spectrum is
dominated by the younger, more luminous stellar population. While the
extremely massive stars are the most luminous, stars with spectral
types O and B (with lifetimes of $<1$ Myr and $\sim 50$ Myr,
respectively) do not survive long. This means that, in a young
population with no ongoing star formation, the stellar light is dominated by that from $A$ stars for
the majority of the first $\sim$ 1 Gyr.

The Balmer absorption lines have become the diagnostic of choice for
measuring the age of the stellar population. For A-type stars or
later, the strength of Balmer absorption lines are good indicators of
stellar temperatures of stars at the main sequence turnoff and,
therefore, good indicators of the main sequence turn-off age. In
addition, the Balmer lines are relatively insensitive to metallicity,
allowing for a strong diagnostic not as confused by the age-metallicity
degeneracy as some other indices.

While we were careful to avoid most of the star-forming disk, in many
cases, our absorption line measurements of H$\beta$ are noticeably
affected by nebular emission fill-in. Due to the superposition of
absorption and emission features, it is difficult to quantify the
degree of contamination from the emission regions.  By comparing data
to models, we are able to determine that the effect is small, in most
cases $< 0.5$ \AA~ for H$\beta$, but it does noticeably affect the
ages that one would infer from the H$\beta$ index. Moreover, in all
cases but one (NGC~4402), we have good measurements of the higher
order Balmer lines (H$\gamma$ and H$\delta$). In fact, it is the ages
inferred from the measurements of these lines that tell us that the
H$\beta$ line is contaminated; in most cases, the age that we would
infer from the H$\beta$ line is {\it older} than the age inferred
from the higher order lines. When comparing the observed spectrum to a
modeled spectrum, we do see a small amount of residual emission in the
H$\alpha$ feature which, assuming a Balmer decrement of
 H$\alpha$/H$\beta$ = 2.86 \citep{osterbrock89}, is enough to account
for the age discrepancy between H$\beta$ and the higher order lines.
Moreover, the age determined from H$\beta$ is much more sensitive to
errors in the measurements than the age estimates from H$\gamma$ or
H$\delta$. Although the errors in H$\gamma$ or H$\delta$ equivalent
widths are higher than those in H$\beta$ by as much as a factor of
3-4, the H$\gamma$ and H$\delta$ models cover $\sim 5$ times the range
in their equivalent widths, compared to H$\beta$. This makes the
errors in the resulting ages somewhat smaller for the higher order
lines. Additionally, because of the high ratio of H$\alpha$ emission
to H$\gamma$ and H$\delta$ emission at (at typical HII region
temperatures of $\sim10^4$K; \citealp{osterbrock89}), there is much
less Balmer emission in the higher order lines. Therefore, while we
present the H$\beta$ models for completeness, we use the average age
concluded from the two higher order Balmer lines as our primary age
diagnostic.

\subsection{Metallicity-Sensitive Index}

As our metallicity-sensitive index, we chose to use the hybrid index
[MgFe]$^\prime$. The index is a weighted average of the Mgb index and
two Fe indices:

\begin{equation}
[\textrm{MgFe}]^\prime \equiv \sqrt{\textrm{Mgb} * (0.72 * \textrm{Fe5270}
  + 0.28 * \textrm{Fe5335})}
\end{equation}

\citet{thomas03} specifically construct this index to be insensitive
to variations in abundance ratios and $\alpha$ element enhancements
for their sample of early type galaxies. While non-solar abundance
ratios typically are not a concern for spiral galaxies, this index is
still useful as an indicator of total metallicity.

\section{Stellar Population Models}
\label{sec-models}

Stellar population models describe how the light from a population of
stars changes as they form and evolve. While models differ in their
details, most share a common, basic method of modeling a stellar
population.  Individual stellar evolution is modeled via isochrones
from a stellar evolution code, a chosen IMF dictates the relative
number of stars, the star formation history parametrizes how those
stars formed as a function of time (i.e. SFR(t)), and empirical
libraries or theoretical stellar atmospheres are used to transform the
theoretical quantities of the isochrones (i.e. $L$ and
$T_{\textrm{eff}}$) into observable quantities (i.e.  broadband colors
or spectra).

\subsection{Starburst99}

For this work, we have chosen to use the Starburst99 (SB99) stellar
population synthesis code \citep{leitherer99}. Many stellar population
synthesis codes use empirical stellar libraries to transform
theoretical parameters into observable quantities (i.e.
\citealp{worthey94,vazdekis99,bc03}). A key shortcoming of many of
these spectral libraries is that they lack young, metal-poor stars
(due to the dearth of these stars in the Solar Neighborhood). The
outer spiral disks of this study contain young stellar
populations. This is one of the primary reasons that we have selected
the SB99 models to compare to the data: the most recent update of the
SB99 models utilizes the synthetic stellar spectra of
\citet{martins05}, as opposed to empirical spectral libraries. These
models include metallicities from Z=0.001 (i.e. 1/20 solar) to Z=0.04
(i.e. $\sim$ twice solar), allowing more complete metallicity
coverage. This means that we avoid the problems of the incomplete
libraries and are able to define a uniform grid of models, from which
we can extract ages and metallicities.

\subsection{Transformation to Observables}

As its output, SB99 produces spectra created by summing several
theoretical spectra of \citet{martins05}. These spectra are then
smoothed to the instrumental resolution of the observed spectra. While
the resolution of the spectra will not affect the equivalent width
measurements, the Lick index measurements can be affected by spectral
resolution mis-match since some line flux can be smoothed into the
pseudo-continuum bandpasses. While this is not ideal as some of the
line flux is shifted out of the line passband, as long as there is
consistency between the observations and models, our conclusions will
be robust. Unlike many stellar population analyses of earlier-type
galaxies, we do {\it not} correct our galaxies for stellar velocity
broadening of absorption lines. Since we are studying the disks of
spirals, the velocity dispersion is very low. In the case where the
effect should be most extreme (NGC~4388, a massive edge-on galaxy), we
find no meaningful difference in results from the corrected and
uncorrected spectra.

\begin{figure}[t]
  \plotone{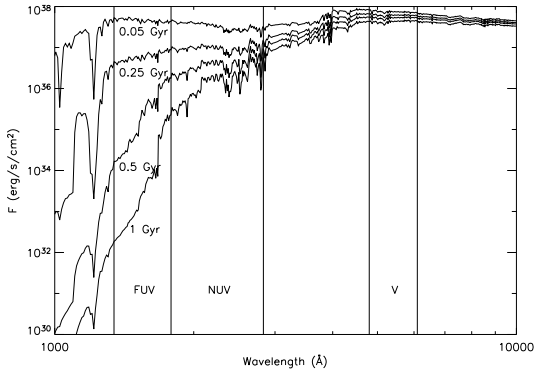}
  \caption[Spectral Evolution]{Evolution of the spectrum from SB99
    quenched star formation models with constant star formation
    followed by instantaneous quenching. Shown are models with quenching times of 50 Myr,
    250 Myr, 500 Myr and 1 Gyr ago. Also indicated on the figure are the
    FUV and NUV bandpasses for GALEX, along with the Johnson broadband
    V bandpass.}
  \label{fig-uvsed}
  \end{figure}

SB99 additionally generates lower resolution SEDs \citep{leitherer99} that extend far
into the UV. These SEDs are used, together with the detector
bandpasses for GALEX to create a grid of colors that depends
critically on age. Specifically, because the FUV flux changes so
quickly at young ages (Figure \ref{fig-uvsed}), the FUV to NUV flux
ratio is a powerful indicator of population age.

\subsection{Modeled Star Formation History}

Much of the stellar population synthesis work in the literature relies
on the ``Simple Stellar Population'' (SSP) assumption; that is, the
assumption that all stars formed at one time and then passively
evolved. While this may be a good approximation of star clusters, it
is clearly {\it not} a good description of the star formation history
in a spiral galaxy; one of the fundamental characteristics of a spiral
is ongoing star formation.  Therefore, we have chosen to model the
stars in the outer disks of these Virgo galaxies as quenched star
formation histories. That is, we assume that there was constant star
formation from a formation time (set to be 12 Gyr ago), followed by rapid quenching of star star formation. We
  define the time since star formation ceased as the quenching time
  ($\tq$). Note that $\tq$ is different from the ``age'' of the
stellar population, which is the value usually quoted when comparing
observations to models. The luminosity-weighted mean age of our
quenched stellar population is significantly {\it larger} than $\tq$.

Practically, the quenched star formation models are constructed by
summing spectra from several bursts. Constant star formation is
approximated as a star forming burst every 50 Myr from the
population's formation time (12 Gyr) to the time of quenching
($\tq$). In addition, we have considered the effect of adding a small
starburst at the time of quenching, in which 2\% of the stellar mass
is formed in a burst at the time of quenching (equivalent to a $\sim
5$ times increase in SFR for 50 Myr). 

The composite theoretical spectra are analyzed in the same manner as
the observed spectra.  LECTOR is used to measure the absorption line
values of each model, and the absorption line strengths extracted from
these spectra allow us to form a line index grid (Figures
\ref{fig-4405sp} - \ref{fig-4424sp}) from which we can extract a value
of $\tq$ for any given galaxy by determining the line of constant age
that passes through the absorption line measurement. The
errors in $\tq$ are then calculated by determining the line of
constant age that pass through the absorption line index $\pm$ the
error in the index.

  \section{The Galaxies}

\label{sec-galaxies}

When we introduce three groups of galaxies in the following section,
we will quote the luminosities and galaxy morphological types (Table
\ref{tab-sample}), in addition to discussing the star-forming
properties. The luminosities (in terms of $L_*$) refer to the $K$ band
luminosities of each galaxy (as determined by 2MASS and assuming a
distance of 16~Mpc). While we list the morphological types, it has
been pointed out that these types do not work well in the Virgo
cluster \citep{kk98}. In particular, nearly all of the galaxies that
we have studied have truncated H$\alpha$ disks. The star formation
properties are drawn from \citet{kk04}, which discusses the star
forming properties for 55 Virgo Cluster spirals. Nearly all of our
galaxies fall into their ``Truncated/Normal'' or ``Truncated/Compact''
categories.  Galaxies classified as ``Truncated/Normal'' have no star
formation beyond a ``truncation radius'' (varies by galaxy between
0.3$R_{25}$ - 0.8$R_{25}$), but normal star formation interior to that
radius.  Galaxies classified as ``Truncated/Compact'' also have an
outer disk free from star formation, but have less regular inner
disks, with non-axisymmetric star formation confined to the inner
$\sim 1$ kpc. We also report the HI deficiency (Def$_{HI}$), as
defined by \citet{gh83} and as measured by
\citet{chung08}. This quantity is the logarithmic ratio of the
HI mass expected for an isolated spiral galaxy to the HI mass measured
in each spiral galaxy. A Def$_{HI}$=0 indicates an HI-normal galaxy,
Def$_{HI}$=1 indicates a galaxy that has 10 times less HI than
expected, and so on. In Table \ref{tab-sample}, we also quote the
projected distance of the galaxies from M87 ($d_{87}$), both in terms
of measured angular separation and computed linear distance (assuming
a distance of 16~Mpc).

For all of the galaxies in our sample, we present a
timescale that represents the time since quenching, assuming a quenched star formation model-- $\tq$. The value of
$\tq$ quoted is the average of the values determined from H$\delta$A
and H$\gamma$A measurements; the age determined from H$\beta$
measurements tends to be unreliable due to some nebular emission
fill-in.

In Figures \ref{fig-4405sp} - \ref{fig-4424sp}, we present the optical
spectrum results for our sample of Virgo galaxies. Each figure
presents the Balmer line strengths (H$\delta$A, H$\gamma$A, and
H$\beta$) plotted against the [MgFe]' strength (these line strengths
are also listed in Table \ref{tab-linestrengths} and the full spectra
can be seen in Figures \ref{fig-gp1sppos} - \ref{fig-gp3sppos}). We
compare these line strengths to Starburst99 models for two different
star formation histories. The top row of the figures shows an
evolutionary track for a SSP model with solar metallicity, with some
sample ages (in Gyr) marked along the line. While the star formation
history of spiral galaxies is likely markedly different from an SSP,
these models are shown here for comparison. Models are displayed in
the middle rows for galaxies with a constant star formation rate to
the quenching time, then no star formation for a range of
metallicities: Z=0.008, Z=0.02 (solar), and Z=0.04. Finally, in the
bottom row, we show the region of the spectra with Balmer line
absorption. In each case, we determine the time since star formation
was quenched by comparing the measured line strengths to SB99 models
(also see Tables \ref{tab-timescales_indiv} \& \ref{tab-timescales}).

\begin{deluxetable*}{c|c|c|c|c|c|c|c|c}
  \tablewidth{0pt}
  \tabletypesize{\small}
  \tablecaption{Properties of Sample Galaxies}
  \tablehead{\colhead{Galaxy} & \colhead{$d_{87}$ (deg)} &
  \colhead{$d_{87}$ (Mpc)} & \colhead{$v_r$ (km/s)} & \colhead{Type} &
  \colhead{SF Class\tablenotemark{\dag}} & \colhead{L
  (L$_{*\mathrm{,K}}$)} & \colhead{$i (\degr)$} &
  \colhead{Def$_{HI}$\tablenotemark{\ddag}}}
\startdata
NGC 4064 & 8.8 & 2.45 & 913 & SBc(s) & T/C & 0.6 & 75 & 1.75\\
NGC 4388 & 1.3 & 0.35 & 2524 & Sb & T/N & 0.9 & 82 & 1.18\\
NGC 4402 & 1.4 & 0.40 & 239 & Sb & T/N & 0.6 & 80 & 0.74\\
NGC 4405 & 4.0 & 1.11 & 1747 & Sc/S0 & T/N [s] & 0.3 & 53 & 0.98\\
NGC 4419 & 2.8 & 0.78 & -196 & Sa & T/A & 0.5 & 82 & 1.23\\
NGC 4424 & 3.1 & 0.87 & 440 & Sa pec & T/C & 0.3 & 63 & 0.75\\
IC 3392 & 2.7 & 0.75 & 1683 & Sc/Sa & T/N [s] & 0.3 & 70 & 1.16\\
NGC 4522 & 3.3 & 0.92 & 2328 & Sc/Sb & T/N [s] & 0.5 & 78 & 0.86\\
NGC 4569 & 1.7 & 0.47 & -220 & Sab(s)I-II & T/N [s] & 3.4 & 66 & 1.52\\
NGC 4580 & 7.2 & 2.0 & 1033 & Sc/Sa & T/N [s] & 0.5 & 42 & 1.44\\
\enddata
\label{tab-sample}
\tablenotetext{\dag}{Star Formation Class, as defined by
  \citet{kk04}: (A) anemic, (T/A) truncated/anemic, (T/C)
  truncated/compact, (T/N) truncated/normal, (T/N [s])
  truncated/normal (severe)}
\tablenotetext{\ddag}{HI Deficiency from \citet{chung08}, as defined by
  \citet{gh83}.}

\end{deluxetable*}

\begin{deluxetable*}{lrrrrrrr}
\tablecolumns{8}
\tablewidth{0pt}
\tabletypesize{\small}
 \tablecaption{Absorption Line Strengths for Composite Spectra}
\tablehead{\colhead{Galaxy} & \colhead{H$\delta$A (\AA)} &
      \colhead{H$\gamma$A (\AA)} & \colhead{H$\beta$ (\AA)} &
    \colhead{Mgb (\AA)} & \colhead{Fe5270 (\AA)}  &  \colhead{Fe5335 (\AA)} & \colhead{[MgFe]$^\prime$ (\AA)}}
  \startdata

  NGC~4064 & $ 2.15 \pm  0.68$ & $-0.66 \pm  0.51$ & $ 2.84 \pm  0.24$ & $ 2.22 \pm  0.23$ & $ 2.45 \pm  0.25$ & $ 2.30 \pm  0.28$ & $ 2.31 \pm  0.15$ \\ 
NGC~4388 & $ 3.39 \pm  0.98$ & $ 1.62 \pm  0.76$ & $ 3.78 \pm  0.40$ & $ 2.33 \pm  0.41$ & $ 2.15 \pm  0.44$ & $ 1.37 \pm  0.50$ & $ 2.12 \pm  0.26$ \\ 
NGC~4402 & \nodata & $ 1.28 \pm  0.77$ & $ 2.57 \pm  0.39$ & $ 1.49 \pm  0.38$ & $ 1.96 \pm  0.40$ & $ 1.11 \pm  0.46$ & $ 1.60 \pm  0.25$ \\ 
NGC~4405 & $ 1.97 \pm  0.80$ & $-1.16 \pm  0.64$ & $ 3.11 \pm  0.31$ & $ 2.49 \pm  0.31$ & $ 2.26 \pm  0.33$ & $ 1.90 \pm  0.37$ & $ 2.32 \pm  0.20$ \\ 
NGC~4419 & $ 1.66 \pm  0.65$ & $-2.94 \pm  0.48$ & $ 1.97 \pm  0.21$ & $ 2.94 \pm  0.19$ & $ 2.45 \pm  0.21$ & $ 2.30 \pm  0.23$ & $ 2.66 \pm  0.13$ \\ 
NGC~4424 & $ 3.20 \pm  0.71$ & $ 0.88 \pm  0.54$ & $ 3.24 \pm  0.27$ & $ 2.08 \pm  0.27$ & $ 2.15 \pm  0.29$ & $ 1.85 \pm  0.32$ & $ 2.07 \pm  0.18$ \\ 
IC~3392 & $ 1.15 \pm  0.80$ & $-1.43 \pm  0.59$ & $ 2.80 \pm  0.28$ & $ 2.34 \pm  0.27$ & $ 2.47 \pm  0.29$ & $ 2.31 \pm  0.32$ & $ 2.38 \pm  0.18$ \\ 
NGC~4522 & $ 5.65 \pm  0.64$ & $ 4.18 \pm  0.51$ & $ 4.33 \pm  0.28$ & $ 1.61 \pm  0.30$ & $ 1.78 \pm  0.32$ & $ 1.58 \pm  0.37$ & $ 1.67 \pm  0.20$ \\ 
NGC~4569 & $ 2.35 \pm  0.68$ & $-1.10 \pm  0.50$ & $ 2.64 \pm  0.23$ & $ 2.80 \pm  0.23$ & $ 2.62 \pm  0.24$ & $ 2.29 \pm  0.27$ & $ 2.66 \pm  0.15$ \\ 
NGC~4580 & $-0.26 \pm  0.86$ & $-3.39 \pm  0.65$ & $ 2.48 \pm  0.29$ & $ 3.23 \pm  0.28$ & $ 2.83 \pm  0.30$ & $ 2.53 \pm  0.33$ & $ 2.97 \pm  0.18$ \\ 

\enddata
\label{tab-linestrengths}
\end{deluxetable*}

\subsection{Truncated ``Normal'' Galaxies}

A majority of the galaxies in our sample have normal stellar disks and
H$\alpha$ disks that are truncated at $\sim0.3 R_{25}$ -- $0.8
R_{25}$. The stellar disks, as traced by the $R$ band light are
generally well-fit by elliptical isophotes and do not show any
apparent stellar asymmetries.

\subsubsection{Old Outer Stellar Disks}

\begin{figure*}
\plotone{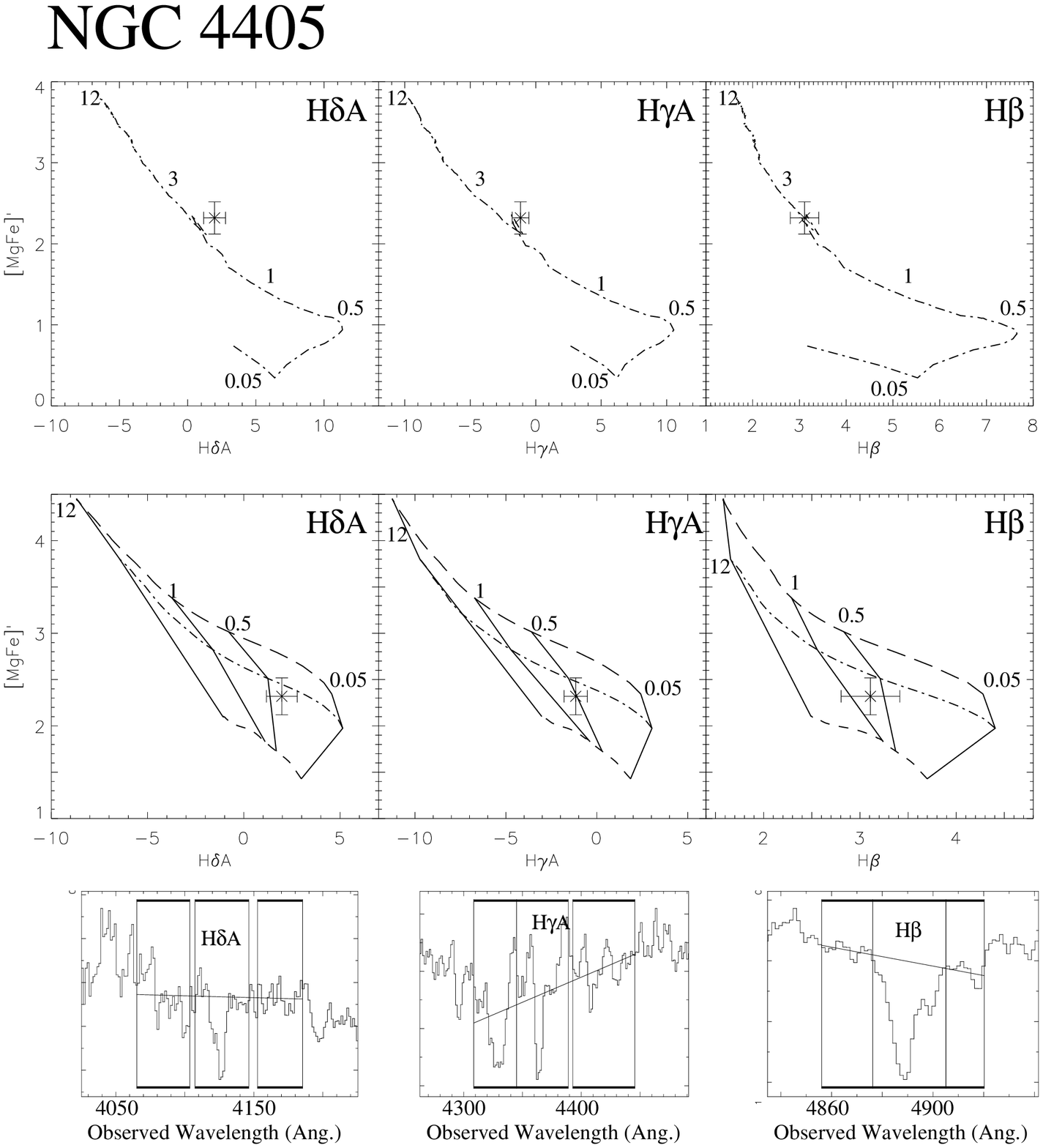}
\caption[NGC 4405 Absorption Index Measurements]{Absorption line indices
  measured for H$\delta$ (left column), H$\gamma$ (middle column), and
  H$\beta$ (right column). The top row shows comparison with a solar
  metallicity SSP, which indicates the luminosity-weighted mean age of
  the population. The middle row shows comparison with a quenched
  model, with $\tq$ indicated by vertical solid lines corresponding
  to 0.05, 0.5, 1.0 and 12.0 Gyr. The three broken lines correspond to
  metallicities of Z=0.008 (short dash), Z=0.02 (i.e. solar;
  dot-dash), and Z=0.04 (long dash). The bottom row shows the region
  of the spectrum from which the Balmer lines were extracted with the
  index bandpass, the continua bandpasses, and the continuum line
  indicated.}
\label{fig-4405sp}
\end{figure*}

\begin{figure*}
\plotone{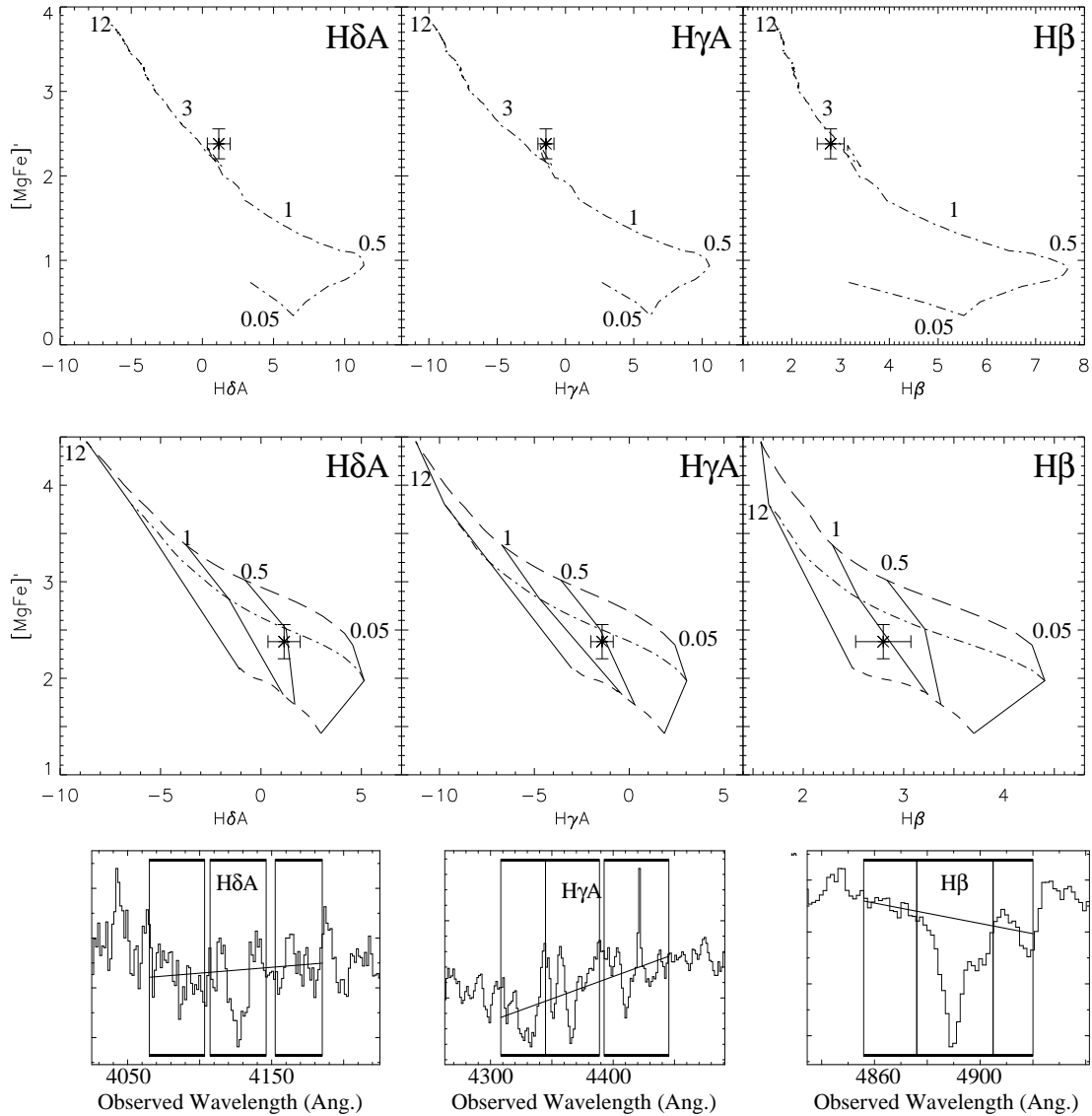}
\caption[IC 3392 Absorption Index Measurements]{Same as Figure
  \ref{fig-4405sp}, but for IC 3392.}
\label{fig-3392sp}
\end{figure*}

\begin{figure*}
\plotone{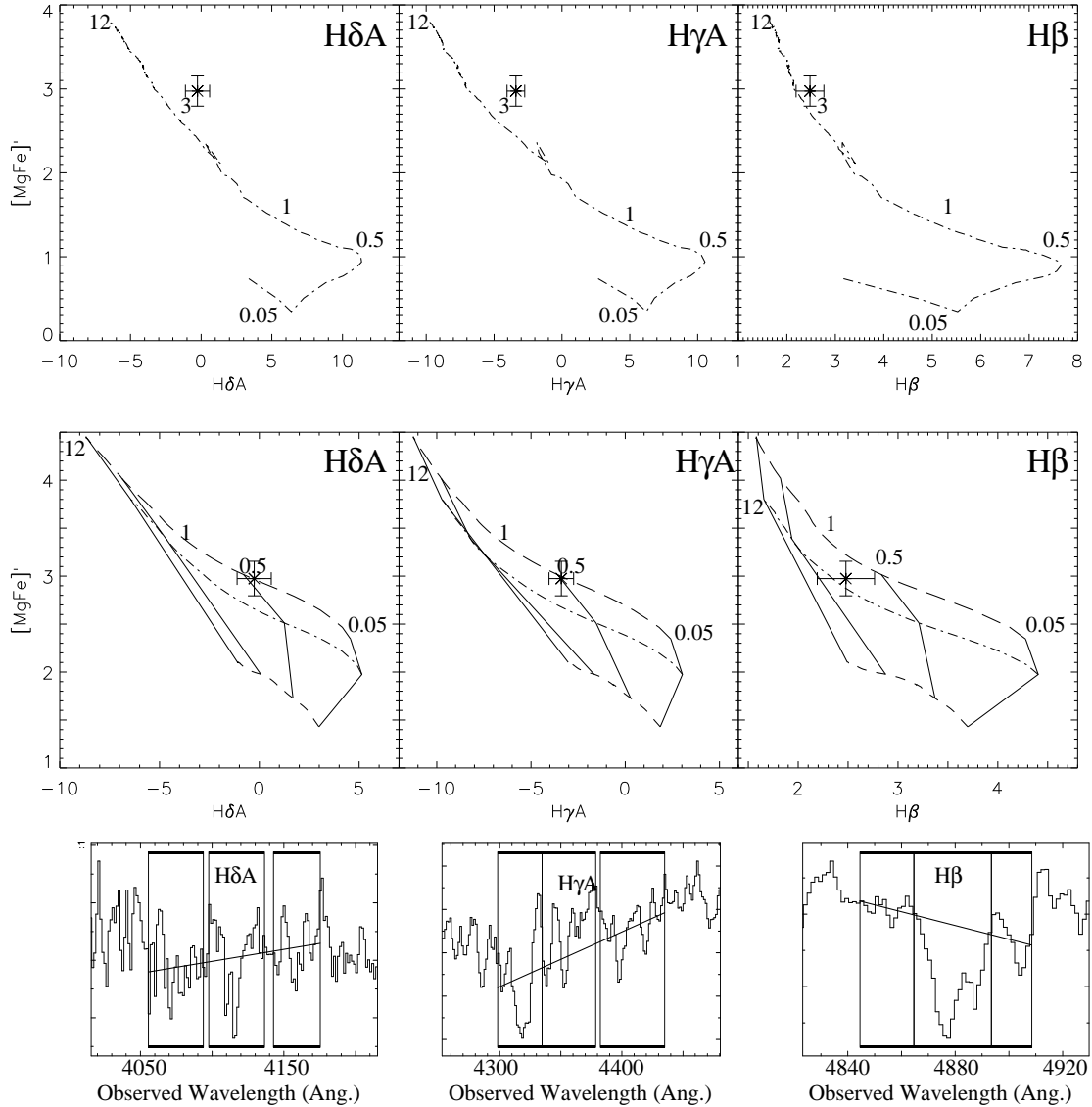}
\caption[NGC 4580 Absorption Index Measurements]{Same as Figure
  \ref{fig-4405sp}, but for NGC 4580.}
\label{fig-4580sp}
\end{figure*}

\begin{figure*}
\plotone{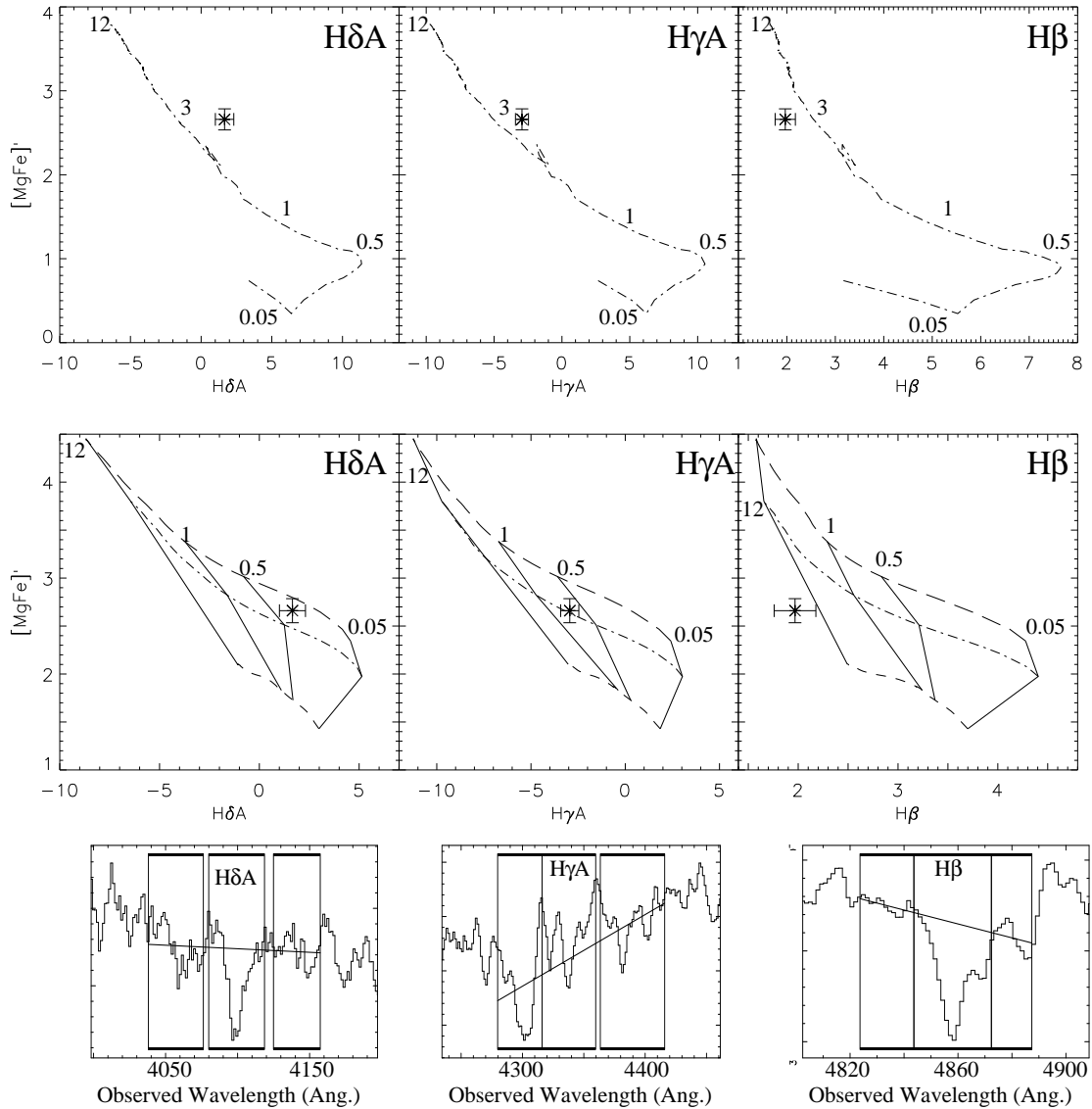}
\caption[NGC 4419 Absorption Index Measurements]{Same as Figure
  \ref{fig-4405sp}, but for NGC 4419.}
\label{fig-4419sp}
\end{figure*}

A subset of the Truncated ``Normal'' Galaxies are those with old outer
stellar disks. The H$\alpha$ morphology for NGC~4405  (Figure
\ref{fig-4405sp}), IC~3392 (Figure \ref{fig-3392sp}),
and NGC~4580 (Figure \ref{fig-4580sp}) are virtually identical; all three galaxies have
H$\alpha$ disks that are small compared to their optical disks and
dramatically truncated at a small fraction of their optical radius.
NGC~4419 (Figure \ref{fig-4419sp}) also has a truncated H$\alpha$ disk, but the existing
H$\alpha$ emission is much more anemic than the other three galaxies.
All four of the galaxies have undisturbed outer stellar disks. The
outer disk of NGC~4580 is somewhat unusual because, despite a lack of
ongoing star formation, there are stellar spiral arms in the outer
disk.

For all four galaxies, the optical spectrum is relatively flat, but
shows a significant break in the blue region, leading to an overall
``red'' color. Comparison of the spectra to models tells us that the
star formation in all of these galaxies was quenched $\sim 450$ Myr
ago or longer. None show any evidence for extraplanar HI or asymmetric
star formation and all are highly HI deficient, with, at most, 10\% of
their ``expected'' HI content \citep{chung08,gh83}. All of these
galaxies have been observed with GALEX and none of them have
significant FUV emission in the outer disk, as would be expected for
regions of galaxies with no star formation within the last $\sim 400$
Myr.

\subsubsection{Young Outer Stellar Disks}

\begin{figure*}
\plotone{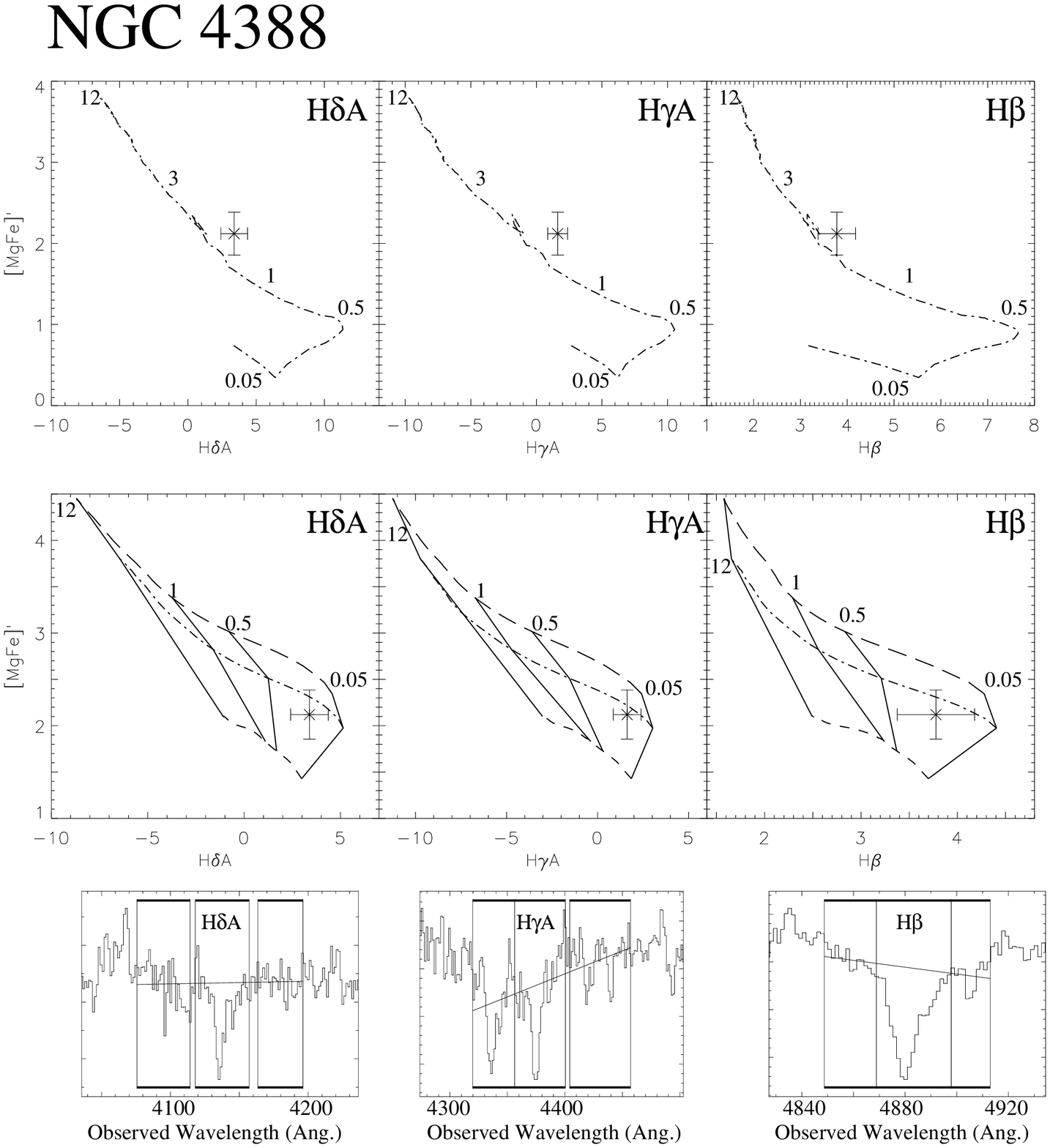}
\caption[NGC 4388 Absorption Index Measurements]{Same as Figure
  \ref{fig-4064sp}, but for NGC 4388.}
\label{fig-4388sp}
\end{figure*}

\begin{figure*}
\plotone{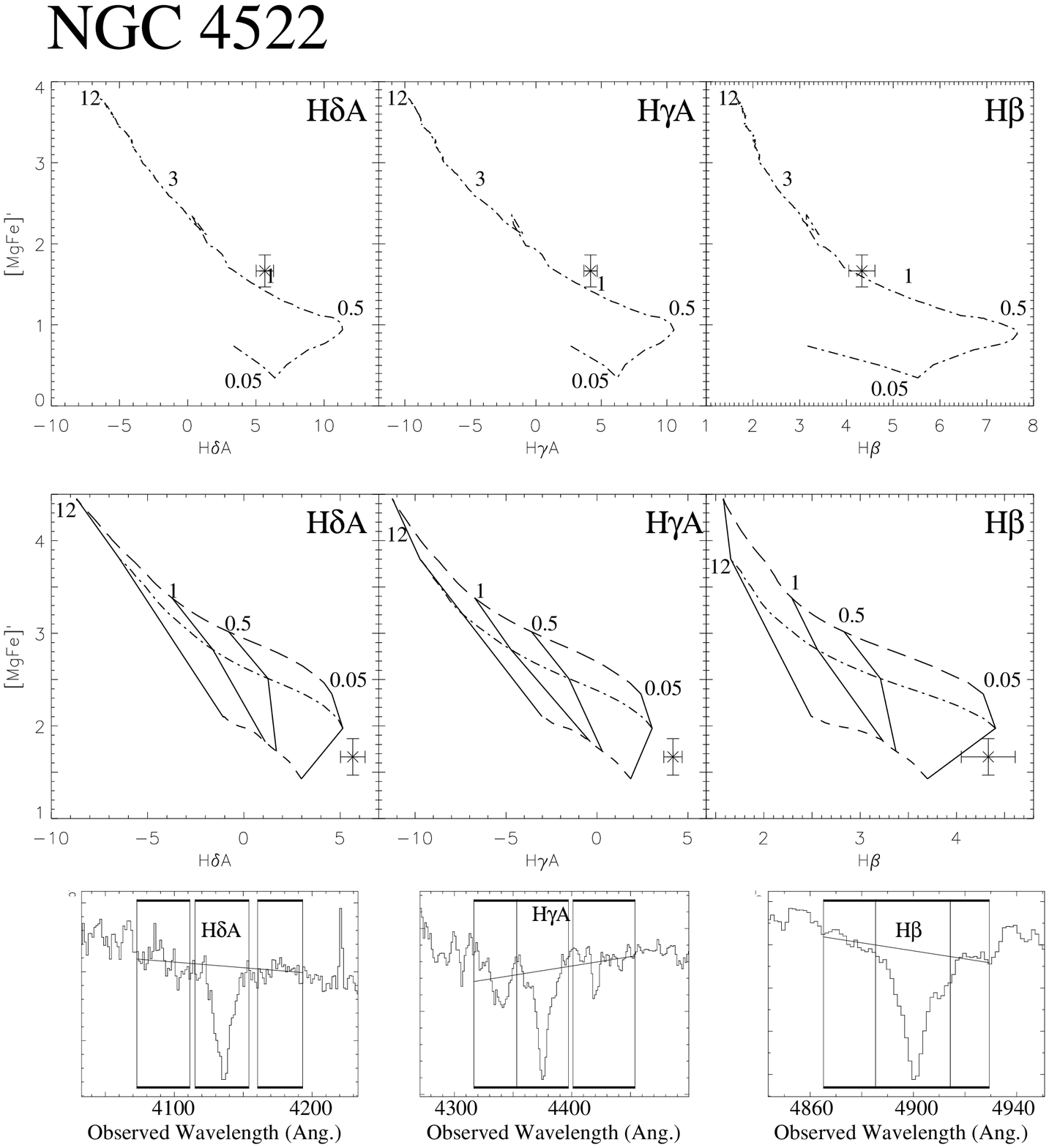}
\caption[NGC 4522 Absorption Index Measurements]{Same as Figure
  \ref{fig-4064sp}, but for NGC 4522.}
\label{fig-4522sp}
\end{figure*}

\begin{figure*}
\plotone{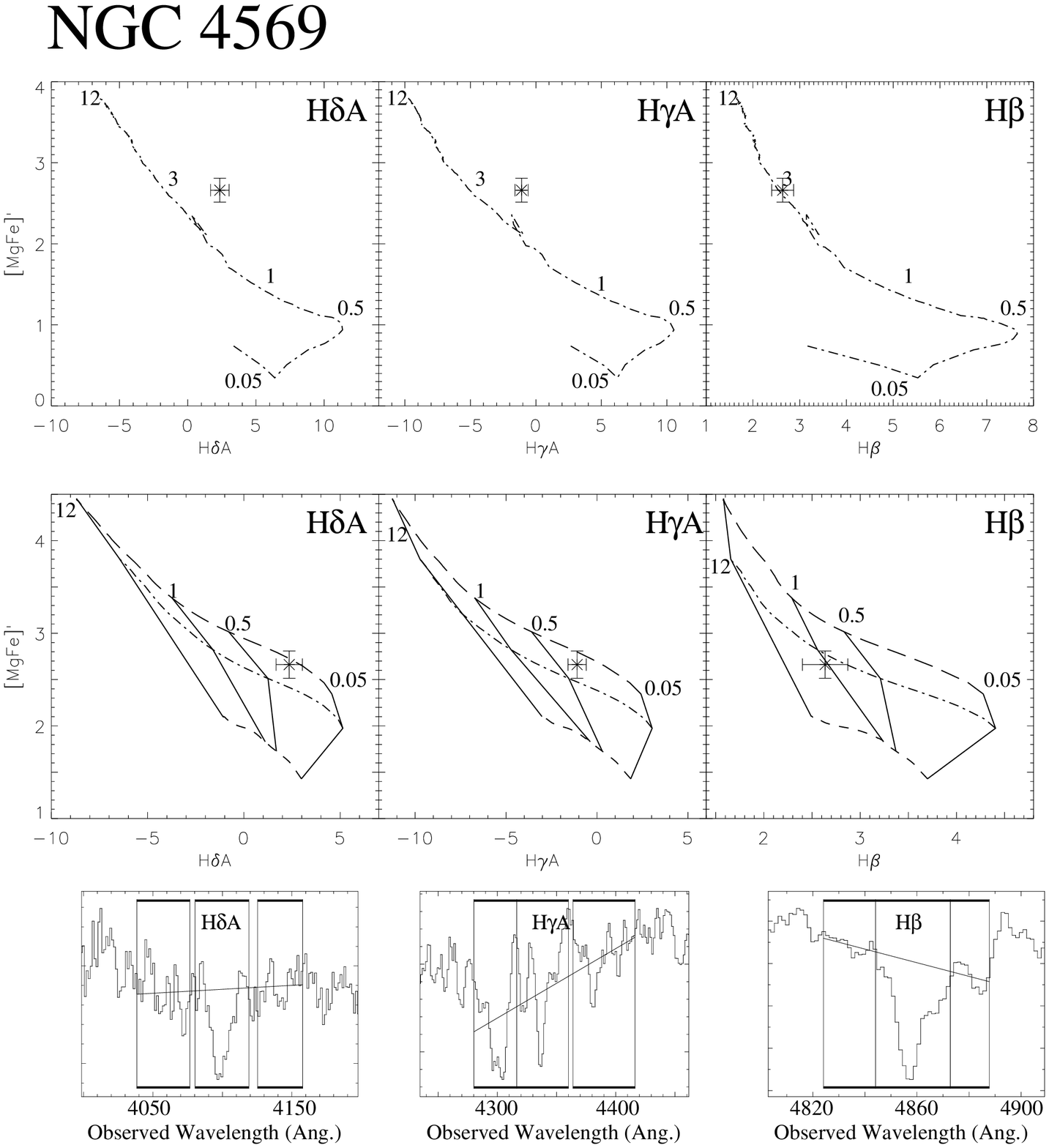}
\caption[NGC 4569 Absorption Index Measurements]{Same as Figure
  \ref{fig-4064sp}, but for NGC 4569.}
\label{fig-4569sp}
\end{figure*}

A second subset of those galaxies with undisturbed stellar disks are
those that have {\it young} outer stellar disks. These include
NGC~4388 (Figure \ref{fig-4388sp}), NGC~4522 (Figure
\ref{fig-4522sp}), and NGC~4569 (Figure \ref{fig-4569sp}). These
galaxies all have disturbed HI spatial distributions
\citep{oosterloo05,kenney04,kenney08}. The star formation in all of
these disks is highly truncated. NGC~4388 has a small, relatively
symmetric H$\alpha$ disk, although there is a known H$\alpha$
\citep{yoshida02} and HI \citep{oosterloo05} tail extending to the
northeast. The star formation in the disk of NGC~4522 is strongly
truncated, although there is significant extraplanar H$\alpha$ and HI
emission to the northwest of this galaxy \citep{kenney04}. NGC~4569
also has strongly truncated H$\alpha$ emission, but has an HI and
H$\alpha$ arm that extends out from the NE edge of the truncated gas
disk (see the H$\alpha$ image in Figure \ref{fig-gp3sppos})
\citep{vollmer04,kenney08}.

The spectra of the outer disks of these galaxies are either moderately
(NGC~4388, NGC~4569) or extremely (NGC~4522) blue, indicating young
stellar populations. All of the galaxies show strong Balmer line
absorption in the outer disk, with NGC~4522 having extremely strong
absorption lines. In addition, all of these galaxies have moderately
weak metal lines when compared to galaxies with older stellar
populations; this is expected in a population with a higher fraction
of A-type stars. In addition, all of these galaxies have significant
GALEX emission, both in the FUV and NUV. NGC~4522 is the extreme case
once again, as the only galaxy outer disk with stronger FUV emission
than NUV emission. By combining UV data and optical spectroscopy, we
conclude that star formation persisted in the stripped disk of
NGC~4522 up to 50-100 Myr ago \citep{crowl06}.

\begin{figure*}
\plotone{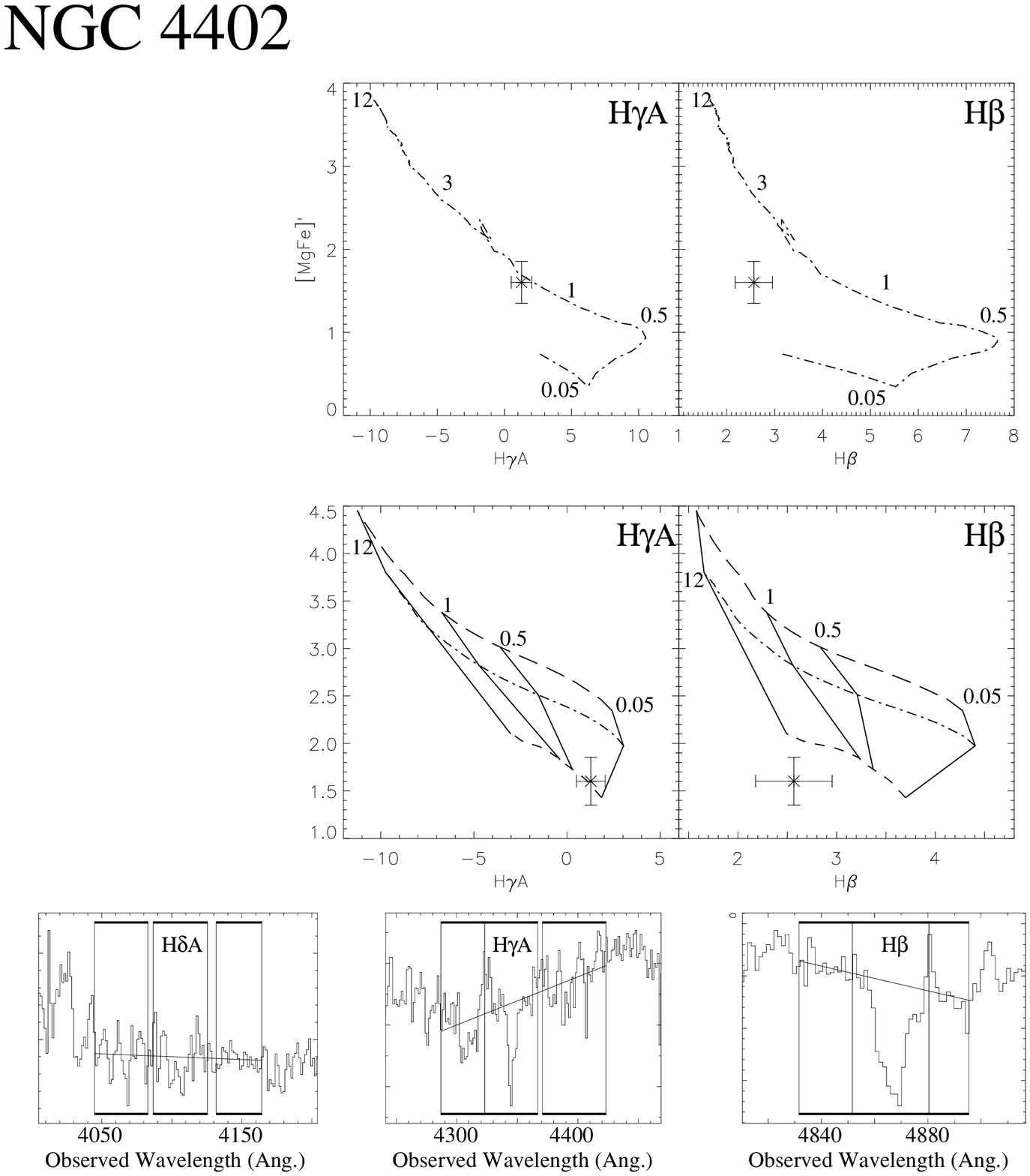}
\caption[NGC 4402 Absorption Index Measurements]{Same as Figure
  \ref{fig-4064sp}, but for NGC 4402. Note that H$\delta$ is not
  detected, so there is no index diagnostic diagram shown.}
\label{fig-4402sp}
\end{figure*}

Star formation extends out to $\sim 0.7 \textrm{R}_{25}$ in the disk
in NGC~4402 (Figure \ref{fig-4402sp}), so the stellar surface
brightness is relatively low beyond the H$\alpha$ truncation
radius. Therefore, the stellar population results are somewhat
inconclusive. The H$\beta$ index seems to indicate a older population,
but other evidence evidence is in conflict with this, perhaps because
the H$\beta$ line is partially filled in with emission. First, the
GALEX photometry suggest that there is a young stellar population in
the outer disk of this galaxy. Secondly, the H$\gamma$ line is much
stronger than would be expected for an old population (Figure
\ref{fig-4402sp}). In addition, there is abundant evidence from
optical imaging and radio mapping that this galaxy is currently
undergoing stripping
\citep{crowl05,vollmer07,murphy08}. Unfortunately, we are unable to
accurately measure the H$\delta$ line in the outer disk of this galaxy
because the surface brightness of the stellar disk at this location is
too low.

\subsection{Galaxies with Complex Interaction Histories}

Finally, there are two galaxies in the sample that appear to
{\it not} have been simply stripped. In the case of NGC~4064,
\citet{cortes06} note that outer isophotes appear undisturbed, but
that there is intense circumnucular star formation in a bar-like
feature and disturbed central dust morphology. \citet{cortes06}
interpret the broadband morphology, H$\alpha$ morphology, and
disturbed gas distribution as a signature that this galaxy experienced
a gravitational interaction, in addition to possibly being stripped of
its gas through an interaction with the ICM. NGC~4424 also has compact
central star formation in a bar-like feature. In contrast to NGC~4064,
the broadband optical appearance of NGC~4424 is also peculiar, with
shell-like features and banana-shaped isophotes, suggesting a major
gravitational disturbance or a merger \citep{kenney96}. Additionally,
the central 2-3 kpc show disturbed dust lanes and infalling molecular
gas \citep{cortes06}.

\begin{figure*}
\plotone{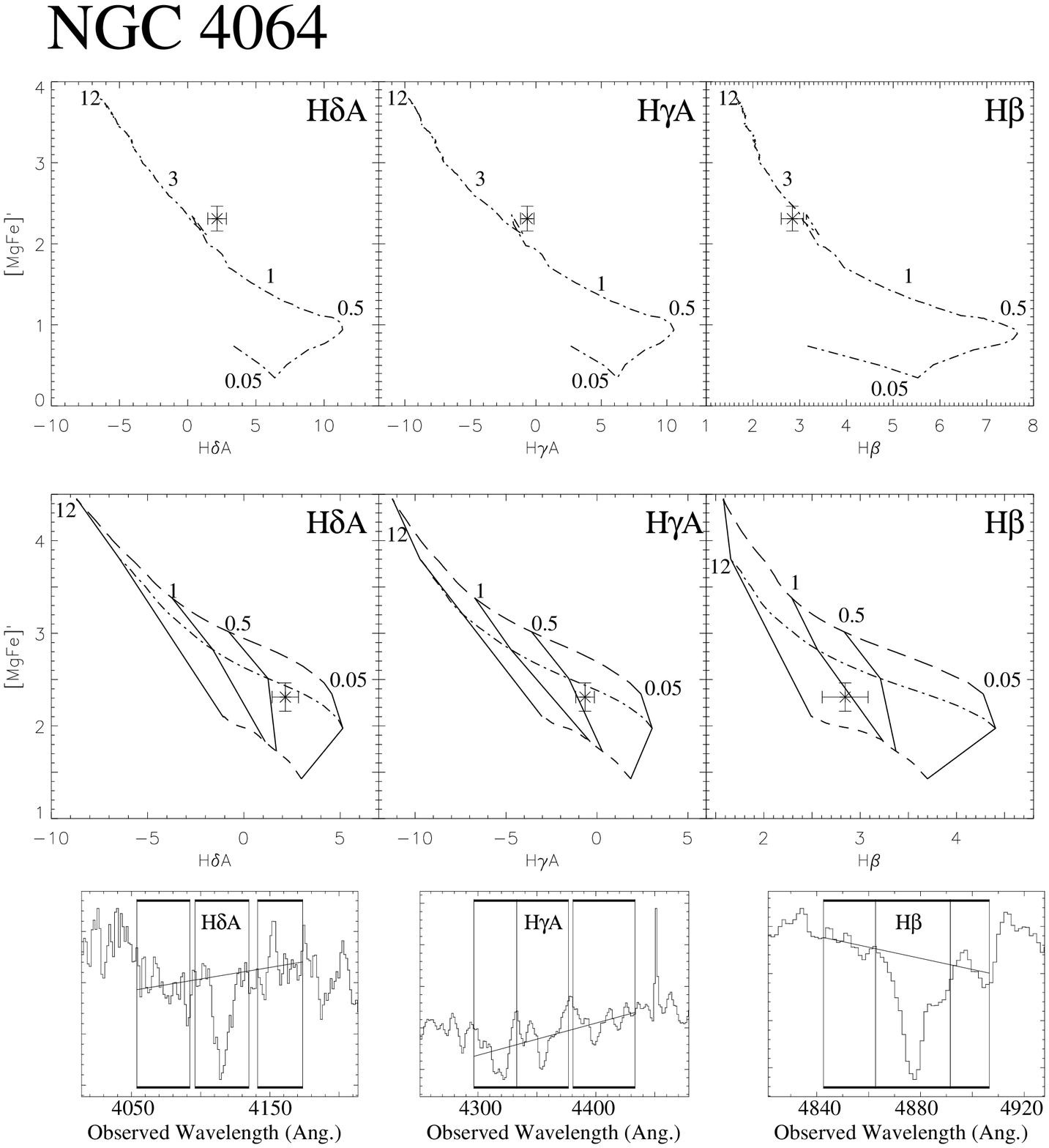}
\caption[NGC 4064 Absorption Index Measurements]{Same as Figure
  \ref{fig-4405sp}, but for NGC 4064.}
\label{fig-4064sp}
  \end{figure*}

The spectrum of NGC~4064 shows characteristics of an intermediate-aged
population (Figure \ref{fig-4064sp}). The spectrum is fairly red and shows strong line
blanketing in the blue part of the spectrum. The metal lines (Mgb and
NaD) are strong, indicative of a population dominated by giant
stars. If we consider the quenching time as extracted from the
H$\delta$ and H$\gamma$ lines, we find $\tq=425\pm75$ Myr. This age
strongly suggests that NGC~4064 was stripped in the cluster outskirts
and not in the core.  See further discussion in Section
\ref{sec-locstrip}.

\begin{figure*}
\plotone{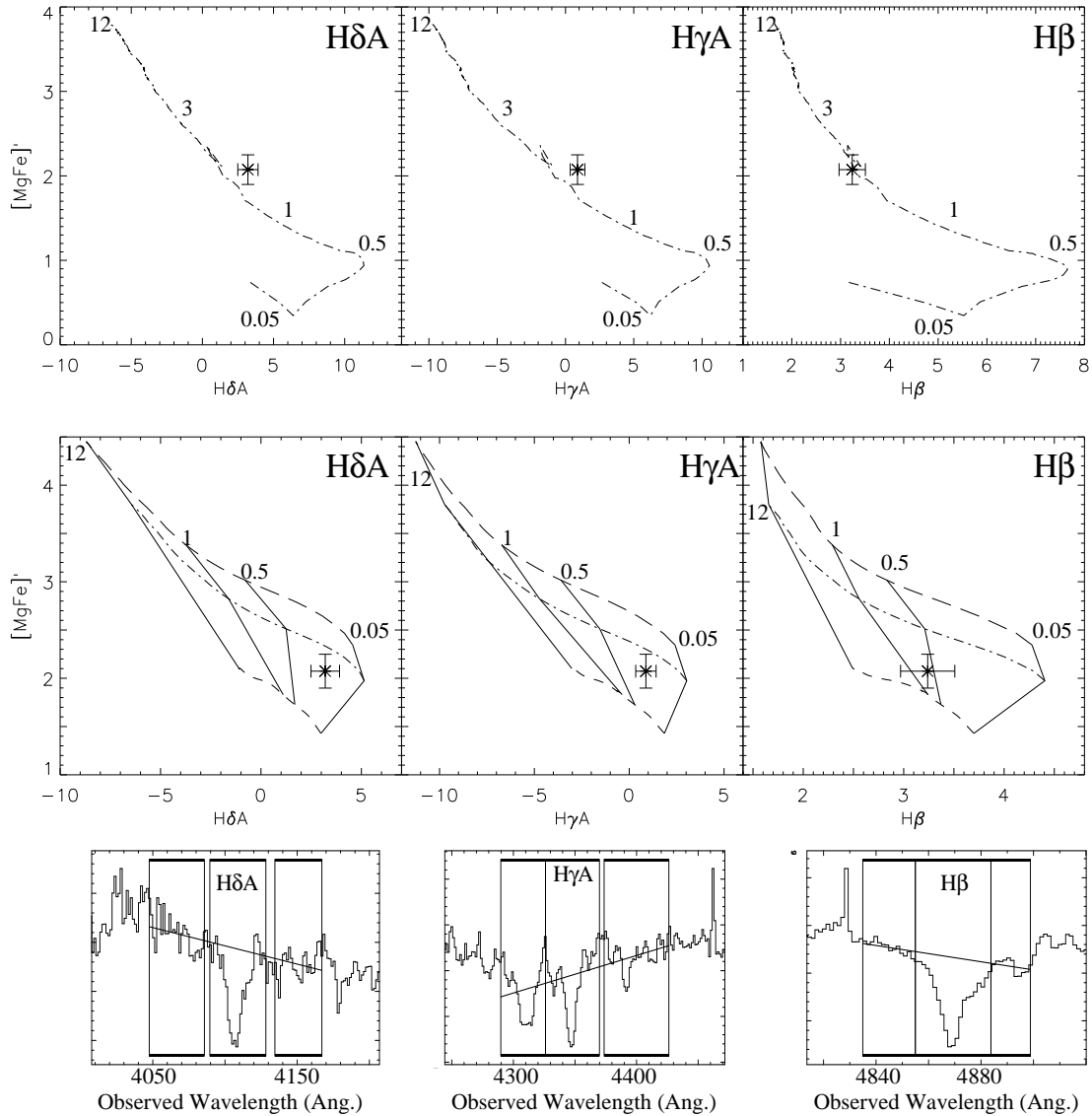}
\caption[NGC 4424 Absorption Index Measurements]{Same as Figure
  \ref{fig-4405sp}, but for NGC 4424.}
\label{fig-4424sp}
\end{figure*}

The stellar population in NGC~4424 is one of the younger in our
sample (Figure \ref{fig-4424sp}). This fact is obvious by looking at the composite spectrum,
which has a moderately blue color, sloping downward to the red part of
the spectrum. Additionally, the Balmer features are quite strong, with
all three major lines easily distinguishable. When comparing these
data to models, we obtain a quenching time of $\tq = 275\pm75$ Myr
for the higher order line indices. From the H$\beta$ measurements, we
infer that there is some modest residual ionized gas emission in the
outer disk, but H$\delta$ and H$\gamma$ give a consistent age, and the
effect appears to be minor.

GALEX observations of NGC~4424 indicate a moderately young disk and
agree with the results from the stellar spectroscopy. The relatively
blue GALEX colors indicate that star formation in this galaxy was quenched $\sim 300$
Myr ago. The timescales from UV imaging and optical spectroscopy only
agree for a quenched star formation history; therefore, these data are
consistent with constant star formation, followed by an instantaneous
quenching.

\begin{deluxetable}{llll}
  \tablewidth{0pt}
  \tablecaption{Time Since End of Star Formation of Quenched Spirals as Determined for
  Individual Balmer Indices}

  \tablehead{\colhead{Galaxy} &
    \colhead{$t_{\mathrm{q,H}\delta\mathrm{A}}$(Gyr)}  &
    \colhead{$t_{\mathrm{q,H}\gamma\mathrm{A}}$(Gyr)}  &
    \colhead{$t_{\mathrm{q,H}\beta}$(Gyr)}
    }
  \startdata
  NGC 4064 &	$0.4\pm0.1$ & $0.45\pm0.1$ & $1.5 \pm 0.9$ \\
NGC 4388 &	$0.25\pm0.1$ & $0.2\pm0.1$ & $0.25 \pm 0.2$ \\
NGC 4402 &	\nodata & $0.2\pm0.15$ 	& $12$ \\
NGC 4405 &	$0.4 \pm 0.15$ 	& $0.5\pm0.15$ & $0.6 \pm 0.2$ \\
NGC 4419 &	$0.35\pm0.1$ & $0.65\pm0.1$ & $>12$ \\
NGC 4424 &	$0.275 \pm 0.075$ & $0.275\pm0.075$ & $0.55 \pm 0.2$ \\
IC 3392	&	$0.5\pm0.1$ & $0.5\pm0.1$ & $1.2_{-0.6}^{+4}$ \\ 
NGC 4522 &	$<0.05$  & $<0.05$ 	& $<0.05$ \\
NGC 4569 &	$0.3\pm0.05$ 	& $0.3\pm0.05$ 	& $1.2_{-0.5}^{+3}$\\
NGC 4580 &	$0.45\pm0.1$ 	& $0.5\pm0.1$ 	& $1.0_{-0.4}^{+3}$\\
  \enddata
  \label{tab-timescales_indiv}
\end{deluxetable}

\begin{deluxetable}{lcl}
  \tablewidth{0pt}
  \tablecaption{Projected Distance from M87 and Derived Stellar
    Population Quenching Timescale for the Stripped Spirals}
  \tablehead{\colhead{Galaxy} & \colhead{d$_{87}$
      (Mpc)} & \colhead{$\tq$ (Myr)}}
    \startdata
  NGC~4064 & 2.45 & $425 \pm 75$ \\
  NGC~4388 & 0.35 & $225 \pm 100$ \\
  NGC~4402 & 0.40 & $<200$ \\
  NGC~4405 & 1.11 & $450 \pm 150$ \\
  NGC~4419 & 0.78 & $500 \pm 150$ \\
  NGC~4424 & 0.87 & $275 \pm 75$  \\
  IC~3392 & 0.75  & $500 \pm 100$  \\
  NGC~4522 & 0.92 & $100 \pm 50$  \\
  NGC~4569 & 0.47 & $300 \pm 50$  \\
  NGC~4580 & 2.01 & $475 \pm 100$  \\
  \enddata
  \label{tab-timescales}
  \end{deluxetable}

  \section{Discussion}
\label{sec-discussion}

\subsection{Ages and Star Formation Histories from GALEX Observations}
\label{sec-galex}

\begin{figure}[t]
\plotone{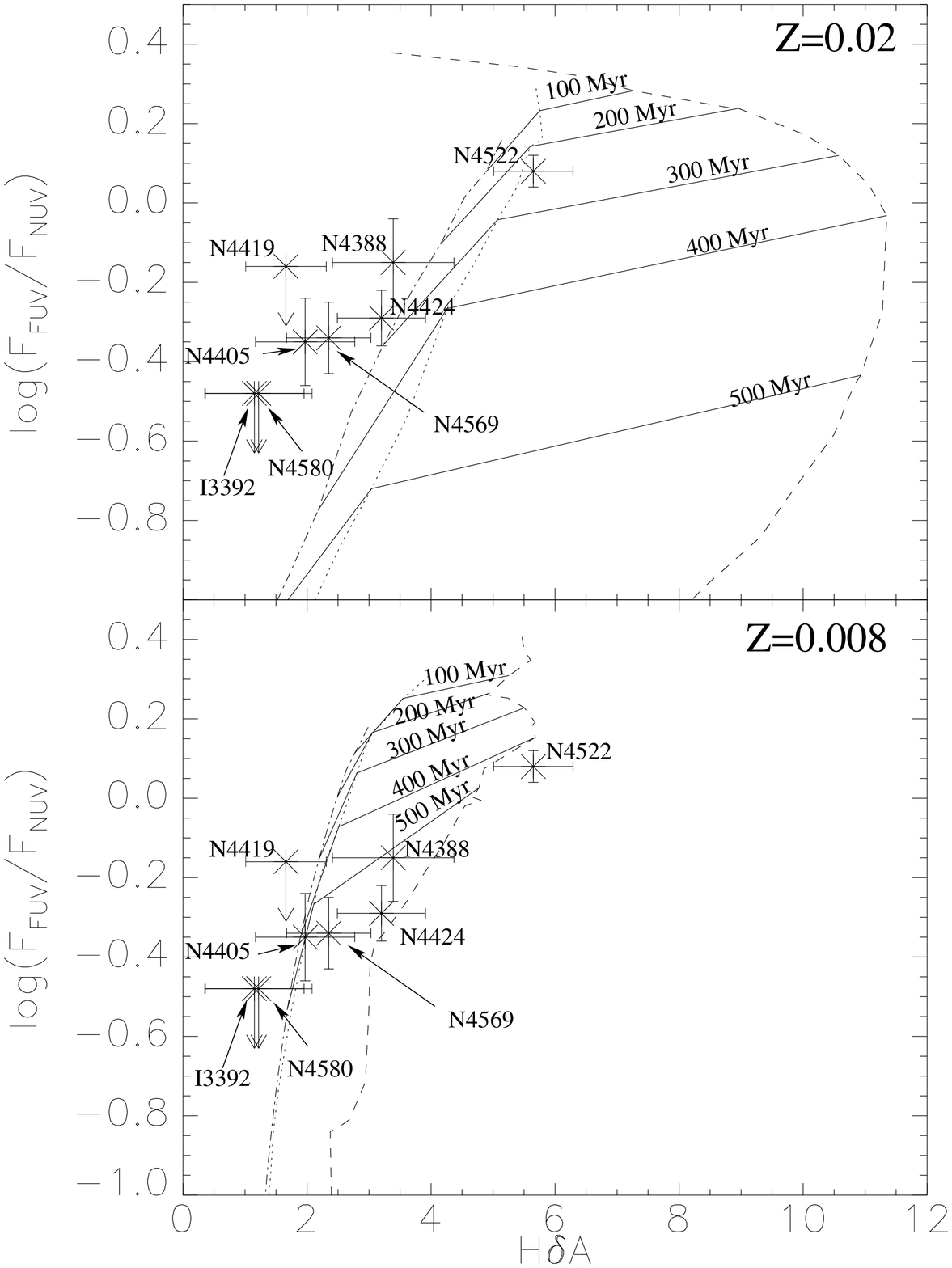}
\caption[GALEX Flux vs. H$\delta$A]{Outer disk H$\delta$A absorption
  lines strengths plotted against the GALEX flux ratio for the same
  outer disk region. Shown as lines on the figure are Z=0.02 (top) and Z=0.008 (bottom) Starburst99 models for a SSP
  (dashed line), a quenched model with a 2\% starburst (dotted line)
  and a quenched model (dash-dot line). The solid lines are
  lines of constant age; from top to bottom: 100, 200, 300, 400, and
  500 Myr.}
\label{fig-galexvshda}
  \end{figure}

  Because of the extraordinary sensitivity of GALEX UV observations to
  the presence of young populations, we can potentially distinguish
  between different star formation histories using our GALEX
  observations.  In particular, we can extract information about a
  galaxy's star formation history by combining analyses of UV
  emission, which traces a very young stellar population and stellar
  absorption lines, which give us information about star formation
  over a longer timescale. There is moderately good agreement between
  the ages determined by spectral analysis of the outer disk and the
  GALEX UV observations (Figure \ref{fig-galexvshda}) when assuming a
  quenching model. There is a clear sequence from NGC~4522, which has
  strong Balmer lines (young ages) as determined by spectral
  measurements and a very hard UV color, to NGC~4580 and IC~3392,
  which have comparatively weak Balmer lines (older ages) and
  significantly softer UV colors. While there is a clear sequence, the
  galaxy measurements are clearly offset from the solar-metallicity
  models (top panel of Figure \ref{fig-galexvshda}). As can be seen
  when comparing to the Z=0.008 models (bottom panel of Figure
  \ref{fig-galexvshda}), this is apparently a metallicity effect,
  showing that there is some degeneracy between star formation
  histories (i.e. quenching, quenching plus a burst, SSP) and
  metallicities. Metallicity estimates from the [MgFe]' index indicate
  that most of the galaxies have a luminosity-weighted mean
  metallicity that is intermediate between Z=0.008 and Z=0.02,
  suggesting that all of the galaxies are consistent with simple
  quenching of star formation, with the exception of NGC~4522 (See
  \citealp{crowl06}).

  As discussed in \citet{crowl06}, the outer disk of NGC~4522 is most
  consistent with a small starburst at the end of its outer-disk star
  formation, such that 2\% of the total stellar mass is formed in a
  very short interval at the time of quenching. As shown in Figure
  \ref{fig-4522sp}, the absorption line indices can not be explained
  with a quenched star formation history; the Balmer lines in the
  outer disk are too strong. It is through GALEX observations (shown
  in Figure \ref{fig-galexvshda} and in \citealp{crowl06}) that we are
  able to obtain a stripping timescale without making an assumption
  about the star formation history.

\subsection{Comparison of Stellar Population Ages to Gas Stripping Simulations}

\begin{figure}[t]
\plotone{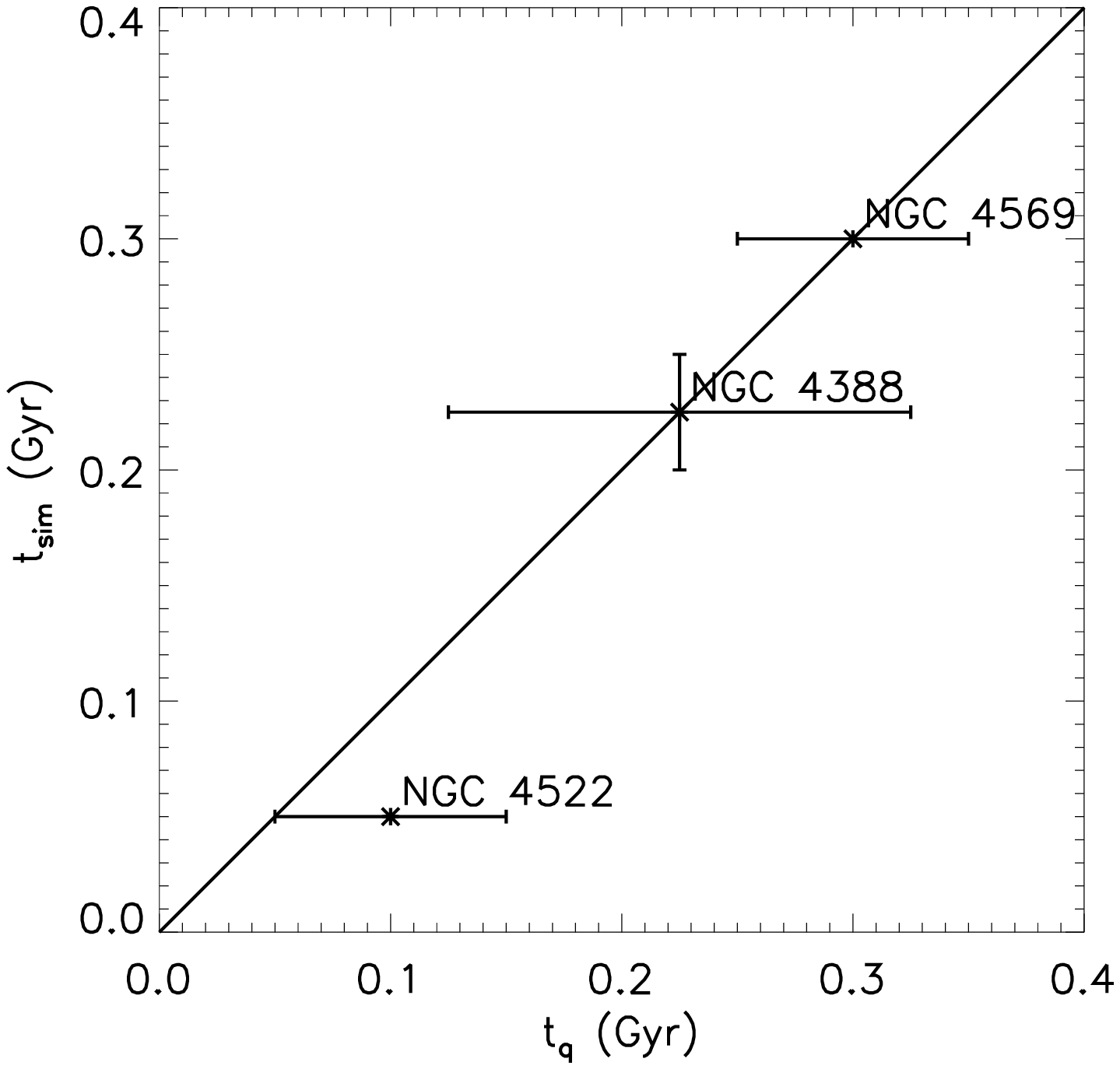}
\caption[$\tq$ vs. Simulation Age]{$\tq$ plotted against the time
  since peak pressure from gas stripping simulations
  ($t_\mathrm{sim}$) for the three galaxies where both data exist. The
  errors for $\tq$ are the errors determined from the stellar
  population models and the error shown in $t_\mathrm{sim}$ for
  NGC~4388 is due to uncertainty in the appropriate model. A 1:1 line
  is shown for comparison.} 
  \end{figure}

\subsubsection{Timescale Comparisons}

For three of the galaxies in our sample, we can compare our timescale
estimates to estimates from a completely independent analysis. While
we observe and model the stellar properties of the stars left behind
after ISM-ICM stripping, it is also possible to measure and model the
gas that has been stripped from these galaxies. For most galaxies in
our sample, the lack of high column density gas near the disk of the
galaxy means it is difficult to constrain the interaction parameters
from gas modeling. However, for three galaxies, NGC~4522, NGC~4388, and
NGC~4569, we are able to compare our results to those from gas
stripping models.

We find a very young star formation quenching age for NGC~4522, $\tq
= 100\pm50$ Myr. This implies that star formation has ended very
recently and, indeed, we do see large amounts of extraplanar gas near
the disk midplane indicating that stripping is ongoing. A comparison
of HI distribution and kinematics with ram-pressure stripping
simulations of this galaxy \citep{vollmer06} imply that peak pressure
was $\sim 50$ Myr ago, in excellent agreement with our results.

For NGC~4388, the stellar population analysis indicates that this galaxy was
stripped longer ago, $\tq = 225$ Myr. A comparison of the HI
distribution and kinematics with a ram pressure simulation of this
galaxy \citep{vollmer03} originally suggested that it was stripped
$\sim 120$ Myr ago. This is in reasonable agreement with our data, but
shows some difference between the HI simulations and the stellar
population ages.  However, recent discoveries of more HI much further
from the galactic plane \citep{oosterloo05} imply that the stripping
must have happened longer ago. Specifically, considering how far the
HI is from the galactic plane, the time since peak pressure should be
revised upward to $\sim$ 200-250 Myr (B. Vollmer, private
communication). While these models are based on much less complete HI
data than the models for NGC~4522 and NGC~4569, the stripping ages
are still in agreement with our estimates from the stellar
population analysis.

For NGC 4569, we find the oldest stripping age of the three: $\tq\sim300$
Myr. This is precisely the age that \citet{vollmer04} derive from
simulations of their HI observations of this galaxy. In fact,
\citet{vollmer04} conclude that the HI features can only be explained
if this galaxy is significantly past the peak pressure phase and much
of the observed HI is falling back onto the disk. Additionally, \citet{boselli06}
model the broadband color profile of NGC~4569 and find that the time
since interaction is constrained to be less than 400 Myr ago. 

\subsubsection{Implications of Timescale Comparisons}

The agreement of our results with those from stripping simulations \citep{vollmer03,vollmer04,vollmer06} is
remarkable for several reasons. First, the agreement of our
  estimates for \textbf{quenching timescales} with their stripping
  timescales is encouraging. While both our
stellar population analysis and HI morphology and kinematic analysis
are built on a large set of assumptions, the assumptions are generally
{\it different} for the two methods. While the stellar population
analysis relies on assumptions about star formation history and
stellar evolution models, the HI simulations rely on how the ICM
interacts with the ISM, how gas is transported away from the disk and
the ram pressure profile that a galaxy experiences while interacting
with the ICM. In the case of these three galaxies, the agreement
between the timescales from HI models and the timescales from our
stellar population analysis suggests that our timescales are probably
correct.

Secondly, the agreement between the ages derived from HI data and
stellar population data suggest that the \textbf{star formation history} we've
assumed is basically correct. While the stellar population models are
largely insensitive to how stars formed early on, if there is a large
burst at the end of star formation, it would lead to artificially
young estimates if modeled with a quenched star formation
history.  In fact, this can be seen for NGC~4522, where 2\% of the
stellar mass has to have formed in a burst at the end of star
formation in order to match the stellar population data. The overall
agreement of ages (even including NGC~4522) indicates that these
galaxies only experience, at most, modest bursts at the end of their
star formation epoch.

Finally, the agreement of our quenching time and the timescales
predicted by gas simulations imply that \textbf{star formation is cut
off when the HI is stripped.} The value of $\tq$ tells us how long
it's been since star formation has ended at the location where we
measure the stellar populations. In the case of our sample of
H$\alpha$ truncated galaxies, we have measured $\tq$ just beyond the
spatial truncation radius. Because of its relative proximity to the
galaxy center, the region just beyond the truncation radius was the
most difficult to strip of any region already stripped by the ICM
interaction. This implies that this region was stripped by the
{\it strongest} pressure the galaxy had yet experienced. In the case
of galaxies that are past peak pressure (as we believe to be the case
for all of our sample), the region just beyond the truncation radius
was stripped when the galaxy was experiencing peak or near-peak ICM
pressure. The agreement between the time elapsed since peak pressure
(from the ram pressure simulations) and the quenching time of star
formation ($\tq$ from our work) means that star formation stops near
the time of peak pressure. This agreement implies that gas stripping
happens quickly and the stripping of the neutral gas (HI) is roughly
coincident with the end of star formation.

\subsection{Locations of Stripping in Virgo}
\label{sec-locstrip}

In a study of nearby clusters, \citet{solanes01} find that, while HI
deficiency extends out to 2$R_A$, the proportion of gas-poor spirals
increases toward the center of the clusters. Additionally,
\citet{gh85} find that the HI deficient galactic fraction correlates
with cluster X-ray luminosity in the sense that those clusters with
higher X-ray luminosities (and presumably higher mass) have more
HI-deficient galaxies.

\begin{figure}[t]
  \plotone{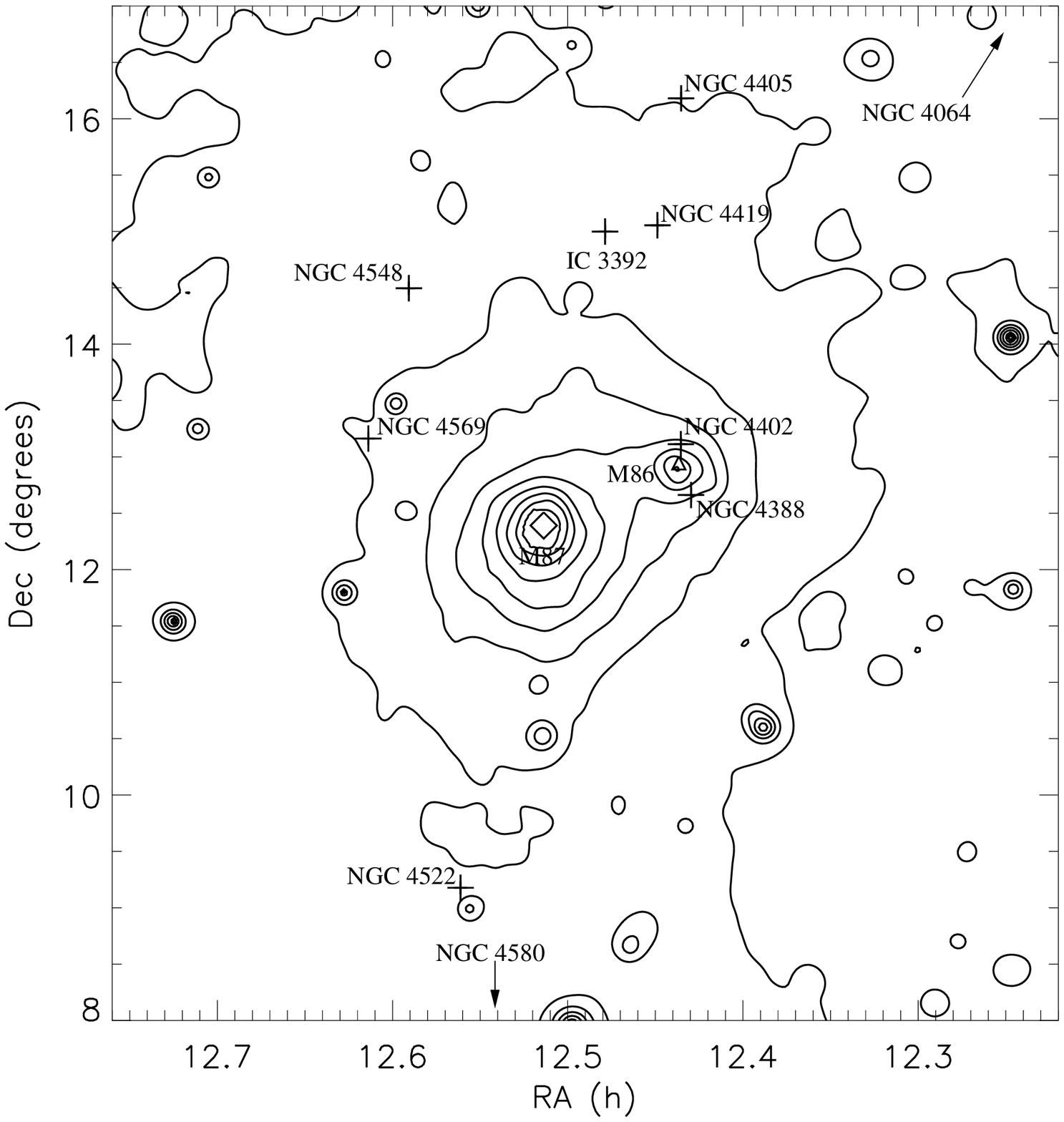}
  \caption[Galaxy Locations Relative to X-Ray Emission]{The positions
  of the sample galaxies compared to locations of hard X-Ray emission
  (0.4-2.4 keV), as traced by images from the ROSAT X-Ray map of Virgo
  Cluster \citep{bohringer94}. Two sample galaxies are not shown here;
  NGC~4064 is off of the figure to the north-west and NGC~4580 is off
  of the figure to the south.}
  \label{fig-contourpos}
\end{figure}

It appears, therefore, that cluster processes are driving HI
deficiency and pushing gas out of galaxies.  If it is the case that
galaxies are only (or primarily) stripped in the core, this implies
that there should be a relationship between galaxies' location in the
cluster (Figure \ref{fig-contourpos}) and time since star formation quenching. While projection
effects limit our understanding of the true distance between a galaxy
and M87 (at the core of Virgo), we can set lower limits to the
clustercentric distance by using the {\it projected} distance from
M87 ($d_{87}$). These are only lower limits because the galaxies are
also displaced from M87 by an unknown amount along the line of sight.

\begin{figure}[t]
  \plotone{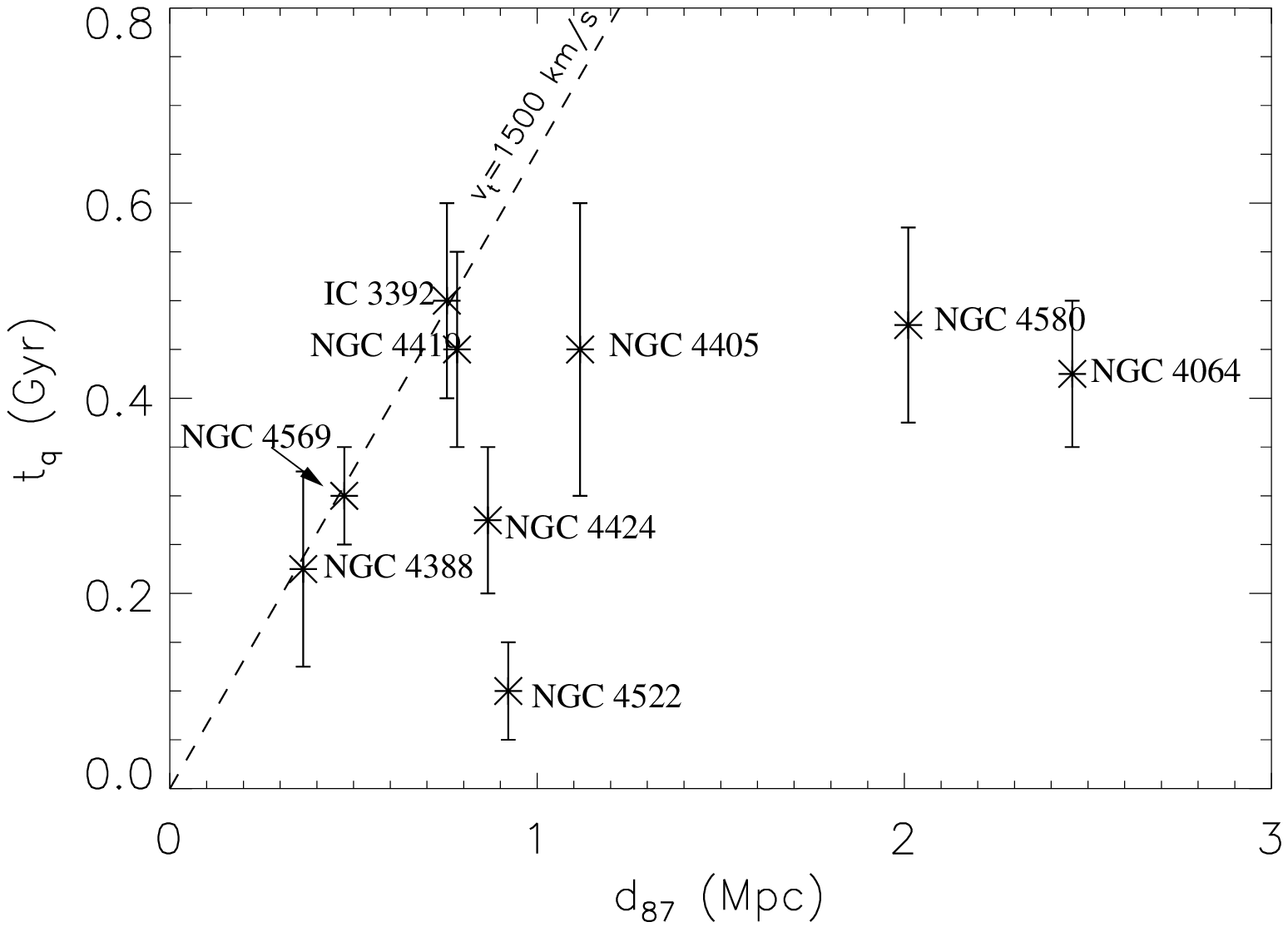}
    \caption[Quenching Time vs. $d_{87}$]{$\tq$ for the sample of
    stripped spirals against projected distance from the central
    elliptical galaxy M87. Also shown (as a dashed line) is the
    position a galaxy would have if its star formation were halted in
    the core and it had been traveling 1500 km/s in the plane of the
    sky away from M87.}
  
  \label{fig-agevsd87}
\end{figure}

In Figure \ref{fig-agevsd87}, we show the distance of the galaxies
from M87 plotted against $\tq$ for each galaxy. The dashed line in
Figure \ref{fig-agevsd87} shows the distance a galaxy could travel in
a given time if its velocity was 1500~km/s in the plane of the
sky. The average line-of-sight velocity of a galaxy in Virgo relative
to the cluster is $\sim 700$km/s, so 1500~km/s is a generous estimate
for most galaxies. Any galaxies to the right of the line in the figure
are either too far out to have been stripped in the cluster core or
have unusually large velocities in the cluster. This is simply
a matter of travel time. Assuming some reasonable velocity, it is
impossible for the galaxies to the right of the line to have reached
their current location in the time since star formation
ended. Therefore, they can not have been stripped in the core. While
several galaxies are consistent with being stripped in the core, there
are three galaxies clearly to the right of the line and one galaxy
marginally to the right of the line; we will take them case by case.

\subsubsection{NGC~4522}

NGC~4522, despite being far from the cluster core, has an exceedingly
young population that precludes it from being stripped in the core. In
addition to its young population, the large amounts of extraplanar HI
and H$\alpha$, and enhanced polarized radio continuum emission and
flat radio spectral index along the leading edge all point to ongoing
strong ram pressure. Simple estimates of ram pressure are too low by a
factor of 10 to adequately explain the observed stripping
\citep{kenney04}. However, the ``simple'' estimates of ram pressure
assume a smooth, static ICM. The existence of galaxies like NGC~4522
suggest that the ICM is either dynamic or ``lumpy.'' Observed ICM
shocks \citep{shibata01} suggest that the ICM may be moving with
significant velocity in the region of NGC~4522.  In \citet{crowl06},
we include a more complete discussion of the stellar populations of
this stripped spiral galaxy.

\subsubsection{NGC~4580}

NGC~4580 is quite far from the cluster core but has an
intermediate-aged outer stellar disk ($\tq = 475$ Myr). In many ways,
NGC~4580 is a prototypical stripped spiral, with normal star formation
within a well-defined radius and no star formation beyond, but with an
undisturbed stellar disk. One slightly peculiar aspect of this galaxy
is the presence of stellar spiral arms beyond the truncation
radius. These stellar spiral arms in the outer disk are stronger than
in any of the other sample galaxies with comparable quenching
times. Without gas, the spiral arms should dissipate relatively
quickly, perhaps within a few galaxy rotation periods, but it appears
these arms have persisted.

\subsubsection{NGC~4064}

NGC~4064 is very far out in the cluster, the furthest from M87 of any
galaxy in our sample. Despite this, its stellar population points to
star formation quenching only 425 Myr ago. It is possible that this
galaxy was stripped of its star forming gas by an interaction with the
ICM, but if that is the case, it did not happen in the core of
Virgo. Optical imaging, kinematics, and CO interferometry
\citep{cortes06} suggest that this galaxy has recently undergone a
minor gravitational interaction. However, this by itself is likely to
not be enough to remove the large amount of gas that this galaxy has
apparently lost. The HI is truncated well within the stellar disk of
NGC~4064, demonstrating that the gas was removed without significantly
disrupting the stars.

\subsubsection{NGC~4424}

NGC~4424 is modestly below the 1500 km/s line and has a
relatively young stripping age. This, coupled with the large HI tail
observed in this galaxy \citep{chung07}, suggests that this galaxy
recently underwent an interaction. Optical morphology and stellar
kinematics \citep{cortes06} demonstrate that NGC~4424 has recently
experienced a significant gravitational interaction. Such interactions
are typically more common in the cluster outskirts, where the relative
velocities of galaxies are lower. However, the large HI tail pointing
toward M49 \citep{chung07} suggests that a collision with that galaxy
may have played a role in the stripping of the HI from NGC~4424. While
this is clearly not the only environmental effect this galaxy has
experienced, it appears that a ram pressure interaction did play a
role in quenching the star formation in NGC~4424.

\subsubsection{Galaxies Stripped in the Core}

The remaining galaxies fall near the $v_t = 1500$ km/s line. This
implies that, within the regions probed by our spectroscopic
observations, star formation was interrupted on a recent core
passage. If the galaxies had been stripped on a previous passage
through the Virgo core, they would fall to the left of the line shown
in the figure; that is, they would be closer to the cluster core than
their stripping age and a travel speed of 1500 km/s would imply.  The
observation that these galaxies are at the minimum age that their
distances allow suggests that the current appearance of these
galaxies, with truncated HI and H$\alpha$ disks, may be a short-lived
state.  One possibility is that these galaxies are stripped further on a
subsequent core passage. Note that because $\tq$ has only been
determined just beyond the truncation radius, we cannot comment on how
much of the outer disk was recently stripped; it's possible that only
a narrow annulus of gas was stripped $\tq$ ago. As a galaxy enters a
cluster, much of its gas is likely stripped on its first core passage
but models suggest that, in a moderate-sized cluster like Virgo, it's
unlikely that {\it all} of the gas will be stripped from an 0.5$L_*$
or 1.0$L_*$ galaxy on that first passage
(e.g. \citealp{vollmer01}). However, on subsequent orbits, gas at
smaller and smaller radii may be stripped. This is particularly likely
if cluster conditions and galaxy orbits change, which are both likely
in an unrelaxed cluster. However, our results appear to show that some
fraction of star-forming gas was truncated on these galaxies' last
passage through the cluster core.

\subsection{Are Stripped Spirals ``Failed'' Passive Spirals?}

With the exception of small star-forming disks in their centers, the
observed stripped spirals in our sample are similar to the ``passive
spirals'' observed at higher redshift \citep{dressler99}. It may be
that Virgo is not massive enough to entirely strip the ISM from the
spiral galaxies we have observed.  However, in the largest clusters,
the ICM density is higher by a factor of $\sim 10$ and the relative
velocities are higher by a factor of $\sim 2$; this means that ram
pressure is higher by a factor of $\sim 40$ in these clusters.
Therefore, it may be that the true passive spirals are only found in
higher mass clusters such as those of \citet{dressler99} and that only
truncated spirals are found in more modest clusters like Virgo.  While
our sample is not a complete sample of truncated spirals in Virgo, it
is still interesting that none of the stripped spirals in our sample
were stripped more than 500 Myr ago. If this observed state (truncated
star-forming disk) were long-lived, we might expect to see stellar
populations with a range of ages; from very newly stripped disks
(i.e. NGC~4522 or NGC~4388) to disks that had been stripped several
gigayears ago. Those galaxies with the most regular, symmetric
H$\alpha$ disks tend to have the oldest values of $\tq$ (i.e.
NGC~4405, IC~3392, NGC~4580), but those values are all less than or
equal to 500~Myr, a relatively short time compared to the orbital time
in the cluster.  We do not know whether the star formation in these
galaxies will fade or the star forming gas will be completely stripped
on a subsequent core passage, but from the stellar populations of the
outer disk, it appears that their current state is not long-lived.

\subsection{Metallicities of Galaxy Populations}

It is generally understood that galaxies create metals through stellar
evolution processes. This means that, broadly, there is an
age-metallicity relationship: in a given galaxy, the stars that form
later tend to be more metal rich than the stars that formed
first. While this is broadly true, the properties of any one region of
a galaxy can be largely determined by the {\it local} star formation
history.  In order to avoid adding free parameters to the SB99 models,
we summed stellar models at a constant metallicity; all stars in a
given model (regardless of the age) have a fixed metallicity. This is,
of course, a gross over-simplification of the stellar population of
the galaxy, but due to the relative insensitivity of the Balmer
indices to metallicity (particularly at the young ages that are most
relevant for our study), we have chosen {\it not} to model any sort
of enrichment history. As a result, the metallicities implied by the
[MgFe]$^\prime$ index are the luminosity-weighted average metallicity
for the entire galaxy and not the metallicity of the current
generation of stars. It should be noted that the luminosity-weighted
average metallicity is slightly sub-solar for most of our sample of
spiral galaxies. This is expected, as galaxies with solar-metallicity
{\it current} generation stars likely have a luminosity-weighted mean
luminosity that is slightly sub-solar; in galaxies with realistic
enrichment histories, the older stellar generations have fewer metals
than the current stellar generation.

\section{Summary}

\label{sec-summary}

We have presented observations of several gas-stripped spiral galaxies
in the Virgo cluster. These are galaxies that have apparently been
partially stripped of their gas by interactions with the ICM, but all
still have ongoing star formation in their centers. We present optical
spectroscopy that demonstrates that the outer disks were stripped of
their star-forming gas within the last 500 Myr.  We find that, while
some galaxies are consistent with being stripped in the cluster core,
others appear to have been stripped outside the core. There is
evidence that some outer cluster galaxies have undergone gravitational
interactions, but others appear consistent with ram pressure
stripping. These galaxies provide evidence that the ICM is not static
and smooth. In at least one case (NGC~4522), there is independent
evidence that the ICM pressure is stronger than simple estimates would
predict. If the effects of ram pressure are important well outside the
cluster core, the HI deficiencies observed at large radii in clusters
\citep{solanes01} can be explained by ISM-ICM interactions. These
interpretations imply that the ``reach'' of the ICM extends well
beyond the cluster core and that ISM-ICM interactions may have a
greater impact on morphologically-driven galaxy evolution than simple
ICM models would suggest.

Furthermore, for three galaxies that have been modeled by gas dynamics
simulations, we find agreement between our determined quenching times
and those simulations. These results imply that our galaxies
experienced essentially constant star formation prior to being
quenched and that there was, at most, a modest starburst at the time
of quenching. It also suggests that star formation is effectively
halted by these strong ICM interactions and molecular gas is
effectively stripped quickly by ICM pressure.

Taken together, our observations show that gas stripping is an
important process in environmentally-driven galaxy evolution. As
ISM-ICM interactions are capable of rapidly halting star formation, it
is reasonable to suppose that a significant passive galaxy population
may evolve from spiral galaxies in clusters. While there may be
multiple paths to S0 galaxies, it appears that, in Virgo, we are
observing a process that can efficiently halt star formation in
cluster spiral galaxies.

\vspace{3mm}

We gratefully acknowledge the advice and assistance of Jim Rose, which
were critical at an early stage of this project. Discussions with
Jacqueline van Gorkom, Aeree Chung, Bernd Vollmer, and Bob Zinn were
also crucial in advancing the science case of our paper. Additionally,
we are grateful to David Schiminovich for providing assistance and
support with the GALEX data, as well as comments on the project and
paper. We gratefully acknowledge NASA's support for construction,
operation, and science analysis for the GALEX mission, developed in
cooperation with CNES of France and the Korean Ministry of Science and
Technology. Finally, we thank the anonymous referee for comments that
improved this paper. This research is supported by NSF Grant
AST-0071251.



\begin{thebibliography}
      
\bibitem[Bershady et al(2004)]{bershady04} Bershady, M.~A., Andersen,
  D.~R., Harker, J., Ramsey, L.~W., \& Verheijen, M.~A.~W.\ 2004,
  \pasp, 116, 565
  
\bibitem[B\"{o}hringer et al.(1994)]{bohringer94} B\"{o}hringer, H.,
Briel, U.~G., Schwarz, R.~A., Voges, W., Hartner, G.,\& Trumper, J. 1994, \nat, 368, 828


\bibitem[Boselli et al.(2006)]{boselli06} Boselli, A., Boissier, 
S., Cortese, L., Gil de Paz, A., Seibert, M., Madore, B.~F., Buat, V., \& 
Martin, D.~C.\ 2006, \apj, 651, 811 



\bibitem[Bruzual \& Charlot(2003)]{bc03} Bruzual, G., \& Charlot, S.\ 
  2003, \mnras, 344, 1000

\bibitem[Christlein \& Zabludoff(2004)]{christlein04} Christlein, D.,
  \& Zabludoff, A.~I.\ 2004, \apj, 616, 192

  \bibitem[Chung et al.(2007)]{chung07} Chung, A., van Gorkom, 
J.~H., Kenney, J.~D.~P., \& Vollmer, B.\ 2007, \apjl, 659, L115

\bibitem[Chung et al. (2008), in prep]{chung08} Chung, A. et
  al. 2008, in prep

\bibitem[Cort{\'e}s et al.(2006)]{cortes06} Cort{\'e}s, J.~R., 
Kenney, J.~D.~P., \& Hardy, E.\ 2006, \aj, 131, 747 

\bibitem[Crowl et al.(2005)]{crowl05} Crowl, H.~H., Kenney, J.~D.~P.,
van Gorkom, J.~H., \& Vollmer, B.\ 2005, \aj, 130, 65

\bibitem[Crowl \& Kenney(2006)]{crowl06} Crowl, H.~H. \& Kenney,
  J.~D.~P.\ 2006, \apjl, 649, L75

\bibitem[Dressler(1980)]{dressler80} Dressler, A.\ 1980, \apj, 
236, 351 


\bibitem[Dressler et al.(1999)]{dressler99} Dressler, A., Smail, 
I., Poggianti, B.~M., Butcher, H., Couch, W.~J., Ellis, R.~S., \& Oemler, 
A.~J.\ 1999, \apjs, 122, 51 

\bibitem[Faber et al.(1985)]{faber85} Faber, S.~M., Friel, 
E.~D., Burstein, D., \& Gaskell, C.~M.\ 1985, \apjs, 57, 711

\bibitem[Giovanelli \& Haynes(1983)]{gh83} Giovanelli, R. \&
    Haynes, M.P. 1983, \aj, 88, 881

\bibitem[Giovanelli \& Haynes(1985)]{gh85} Giovanelli, R., 
\& Haynes, M.~P.\ 1985, \apj, 292, 404

\bibitem[Kauffmann et al.(2004)]{kauffmann04} Kauffmann, G., White, 
S.~D.~M., Heckman, T.~M., M{\'e}nard, B., Brinchmann, J., Charlot, S., 
Tremonti, C., \& Brinkmann, J.\ 2004, \mnras, 353, 713 

\bibitem[Kenney et al.(1996)]{kenney96} Kenney, J.~D.~P., 
Koopmann, R.~A., Rubin, V.~C., \& Young, J.~S.\ 1996, \aj, 111, 152 

\bibitem[Kenney et al.(2004)]{kenney04} Kenney, J.~D.~P., van 
Gorkom, J.~H., \& Vollmer, B.\ 2004, \aj, 127, 3361

\bibitem[Kenney et al.(2008)]{kenney08} Kenney, J.~D.~P. et al. 2008,
  in prep.

\bibitem[Koopmann \& Kenney(1998)]{kk98} Koopmann, R.~A., \& 
Kenney, J.~D.~P.\ 1998, \apjl, 497, L75 

\bibitem[Koopmann, Kenney, \& Young (2001)]{koopmann01} Koopmann, R.~A., 
Kenney, J.~D.~P., \& Young, J.\ 2001, \apjs, 135, 125

\bibitem[Koopmann \& Kenney(2004)]{kk04} Koopmann, R.~A., \& 
Kenney, J.~D.~P.\ 2004, \apj, 613, 866 

\bibitem[Leitherer et al.(1999)]{leitherer99} Leitherer, C., et al.\ 
  1999, \apjs, 123, 3

\bibitem[Li \& Draine(2001)]{li01} Li, A., \& Draine, B.~T.\ 2001,
  \apj, 554, 778
  
\bibitem[Martin et al.(2005)]{martin05} Martin, D.~C., et
  al.\ 2005, \apjl, 619, L1

  \bibitem[Martins et al.(2005)]{martins05} Martins, L.~P., Delgado,
  R.~M.~G., Leitherer, C., Cervi{\~n}o, M., \& Hauschildt, P.\ 2005,
  \mnras, 358, 49
  
\bibitem[Moran et al.(2006)]{moran06} Moran, S.~M., Ellis, 
R.~S., Treu, T., Salim, S., Rich, R.~M., Smith, G.~P., \& Kneib, J.-P.\ 
2006, \apjl, 641, L97 

\bibitem[Morrissey et al.(2005)]{morrissey05} Morrissey, P., et 
al.\ 2005, \apjl, 619, L7 

\bibitem[Murphy et al. (2008)]{murphy08} Murphy, E., Kenney, J.D.P.,
  Helou, G., Chung, A., \& Howell, J.H. 2008, ApJ, submitted

\bibitem[Oosterloo \& van Gorkom(2005)]{oosterloo05} Oosterloo, T., 
\& van Gorkom, J.\ 2005, \aap, 437, L19

\bibitem[Osterbrock(1989)]{osterbrock89} Osterbrock,
D.E. \underline{Astrophysics of Planetary Nebulae and Active Galactic
Nuclei}, University Science Books, 1989

\bibitem[Poggianti et al.(2006)]{poggianti06} Poggianti, B.~M., et 
al.\ 2006, \apj, 642, 188 

\bibitem[Schlegel et al.(1998)]{schlegel98} Schlegel, D.~J., 
Finkbeiner, D.~P., \& Davis, M.\ 1998, \apj, 500, 525 


\bibitem[Shibata et al.(2001)]{shibata01} Shibata, R., 
Matsushita, K., Yamasaki, N.~Y., Ohashi, T., Ishida, M., Kikuchi, K., 
B{\"o}hringer, H., \& Matsumoto, H.\ 2001, \apj, 549, 228 

\bibitem[Shioya et al.(2004)]{shioya04} Shioya, Y., Bekki, K., 
\& Couch, W.~J.\ 2004, \apj, 601, 654 

\bibitem[Solanes et al.(2001)]{solanes01} Solanes, J.~M., 
Manrique, A., Garcia-G{\'o}mez, C., Gonz{\'a}lez-Casado, G., 
Giovanelli, R., \& Haynes, M.~P.\ 2001, \apj, 548, 97 

\bibitem[Thomas et al.(2003)]{thomas03} Thomas, D., Maraston, 
C., \& Bender, R.\ 2003, \mnras, 339, 897 

\bibitem[Vazdekis(1999)]{vazdekis99} Vazdekis, A.\ 1999, \apj, 513,
  224

\bibitem[Vollmer et al.(2001)]{vollmer01} Vollmer, B., Cayatte, 
V., Balkowski, C., \& Duschl, W.~J.\ 2001, \apj, 561, 708 


\bibitem[Vollmer \& Huchtmeier(2003)]{vollmer03} Vollmer, B., \& 
Huchtmeier, W.\ 2003, \aap, 406, 427 

\bibitem[Vollmer et al.(2004)]{vollmer04} Vollmer, B., Balkowski, 
C., Cayatte, V., van Driel, W., \& Huchtmeier, W.\ 2004, \aap, 419, 35 

\bibitem[Vollmer et al.(2006)]{vollmer06} Vollmer, B., Soida, M., 
Otmianowska-Mazur, K., Kenney, J.~D.~P., van Gorkom, J.~H., \& Beck, R.\ 
2006, \aap, 453, 883

\bibitem[Vollmer et al.(2007)]{vollmer07} Vollmer, B., Soida, M.,
Beck, R., Urbanik, M., Chy{\.z}y, K.~T., Otmianowska-Mazur, K.,
Kenney, J.~D.~P., \& van Gorkom, J.~H.\ 2007, \aap, 464, L37
  
\bibitem[Worthey(1994)]{worthey94} Worthey, G.\ 1994, \apjs, 95, 107 

\bibitem[Worthey et al.(1994)]{worthey94b} Worthey, G., Faber, 
S.~M., Gonzalez, J.~J., \& Burstein, D.\ 1994, \apjs, 94, 687 


  
\bibitem[Worthey \& Ottaviani(1997)]{worthey97} Worthey, G., \& Ottaviani, D.~L.\ 1997, \apjs, 111, 377 



\bibitem[Yoshida et al.(2002)]{yoshida02} Yoshida, M., et al.\ 2002,
  \apj, 567, 118


\end{thebibliography}
\end{document}